\documentclass[10pt,twocolumn,twoside]{IEEEtran}

\usepackage{CJK,indentfirst}
\usepackage[compress]{cite}
\makeatletter
\renewcommand{\citepunct}{,\penalty\@m\hskip.13emplus.1emminus.1em}
\renewcommand{\citedash}{\hbox{--}\penalty\@m}
\makeatother
\usepackage{graphicx}
\usepackage{amsmath}
\usepackage{amssymb}
\usepackage{amsfonts}
\usepackage{psfrag}
\usepackage{array}
\ifCLASSOPTIONcompsoc
\usepackage[tight,normalsize,sf,SF]{subfigure}
\else
\usepackage[tight,footnotesize]{subfigure}
\fi

\usepackage[named]{algo}
\usepackage{algorithmic}
\usepackage{nccmath}
\usepackage{threeparttable}
\usepackage{booktabs}

\newcommand{\bs}{\text{PBS}}
\newcommand{\ms}{\text{PUE}}

\IEEEoverridecommandlockouts

\begin{document}
%\begin{CJK*}{GBK}{song}
%\CJKindent

\title{Full Duplex Assisted Inter-cell Interference Cancellation in~Heterogeneous~Networks}


\author{Shengqian Han, \IEEEmembership{Member, IEEE,} Chenyang Yang, \IEEEmembership{Senior Member, IEEE,} and Pan Chen
\thanks{
This work was supported in part by the National Natural Science Foundation of China (No.
61301084) and by the Fundamental Research Funds for the Central Universities.

S. Han, C. Yang, and P. Chen are with the School of Electronics and
Information Engineering, Beihang University, Beijing, China (e-mail:
\{sqhan, cyyang, chenpan\}@buaa.edu.cn).}
}

% mane the title area
\maketitle

\begin{abstract}
The paper studies the suppression of cross-tier inter-cell interference (ICI) generated by a macro base station (MBS) to pico user equipments (PUEs) in heterogeneous networks (HetNets). Different from existing ICI avoidance schemes such as enhanced ICI cancellation (eICIC) and coordinated beamforming, which generally operate at the MBS, we propose a full duplex (FD) assisted ICI cancellation (fICIC) scheme, which can operate at each pico BS (PBS) individually and is transparent to the MBS. The basic idea of the fICIC is to apply FD technique at the PBS such that the PBS can send the desired signals and forward the listened cross-tier ICI simultaneously to PUEs. We first consider the narrowband single-user case, where the MBS serves a single macro UE and each PBS serves a single PUE. We obtain the closed-form solution of the optimal fICIC scheme, and analyze its asymptotical performance in ICI-dominated scenario. We then investigate the general narrowband multi-user case, where both MBS and PBSs serve multiple UEs. We devise a low-complexity algorithm to optimize the fICIC aimed at maximizing the downlink sum rate of the PUEs subject to user fairness constraint. Finally, the generalization of the fICIC to wideband systems is investigated. Simulations validate the analytical results and demonstrate the advantages of the fICIC on mitigating cross-tier ICI.
\end{abstract}
\begin{IEEEkeywords}
Inter-cell interference cancellation, Full duplex, Heterogeneous networks, eICIC, CoMP, OFDM.
\end{IEEEkeywords}
\section{Introduction}

Cellular systems are evolving toward heterogeneous networks (HetNets) with universal frequency reuse in order to support the galloping demand of mobile wireless services \cite{Soret2013}. By deploying low-power nodes such as micro, pico, or femto base stations (BSs) within the coverage of traditional macro BSs (MBSs), HetNets bring the network close to the user equipments (UEs) to obtain the ``cell-splitting'' gain, and therefore improve the area spectral efficiency. For simplicity, we refer to the low-power nodes as pico BSs (PBSs)~hereinafter.

In practice, however, straightforwardly deploying PBSs in the coverage of MBSs cannot effectively realize the promised benefits of HetNets because the large difference of the two types of BSs in transmit power makes pico UEs (PUEs) suffer from severe cross-tier inter-cell interference (ICI) generated by the MBS. Efficient ICI cancellation (ICIC) mechanisms are therefore critical to support the HetNet deployment \cite{Soret2013}.

To mitigate the ICI, enhanced ICIC (eICIC) techniques have been developed in Release~10 of Long-term Evolution (LTE)\cite{Soret2013}. In the time-domain eICIC, the MBS remains silent in the so-called almost blank subframes (ABS), during which the cell-edge PUEs are served without interference. In the frequency-domain eICIC, the MBS and PBSs schedule UEs in orthogonal frequency resources to avoid the ICI. The eICIC methods are of low complexity and easy to implement, but they limit the performance of both macro UEs (MUEs) and PUEs since the UEs can be only served in partial time-frequency resources.
Coordinated multi-point (CoMP) transmission is another promising technique for cross-tier ICI suppression, which exploits the antenna resources of the MBS~\cite{Yang2013}. Considering the fact that providing high-performance backhaul from all PBSs to the core network may be cost prohibitive, coordinated beamforming (CB) based CoMP transmission, requiring no data sharing among the BSs, has received wide attention. CoMP-CB can be considered as a sort of spatial-domain eICIC, which enables the PUEs to be served with the whole time-frequency resources. However, the performance of CoMP-CB is limited by the number of antennas at the MBS, which is usually not acceptable when the PBSs are densely deployed in the coverage of the MBS.

Different from eICIC and CoMP-CB, both of which control the transmission of the MBS in time, frequency or spatial domain in order to generate an ICI-free environment for the transmission between PBSs and PUEs, in this paper we strive to study the cross-tier ICI suppression scheme without the participation of MBSs, which therefore will not consume the resources of the MBS. Specifically, we consider using full duplex (FD) technique to HetNets. FD communication was long believed impossible in wireless system design due to the severe self-interference within the same transceiver. However, the belief has been overturned recently with the tremendous progress in self-interference cancellation~\cite{Choi2010, Duarte2012}, where the plausibility of FD technique for short-range point-to-point communications was approved. FD techniques have been applied to provide bi-directional communications over the same time and frequency resources, and exhibited noticeable spectral efficiency gains over half-duplex (HD) schemes~\cite{Hong2014, Everett2014}. FD techniques are also applied in relay systems to improve service coverage, where the usage of FD avoids the waste of resources as in HD relay systems~\cite{Riihonen2009, Day2012}. To the best of our knowledge, leveraging FD in HetNets to suppress the cross-tier ICI has not been addressed in the literature.

\begin{table}\centering
  \renewcommand{\arraystretch}{0.95}
    \begin{threeparttable}[b]
    \caption{List of major symbols} \label{T:Acronyms}
    \begin{tabular}{l|p{0.31\textwidth}}
        % \toprule
        \hline
        $B, K_M, K_P$  & Numbers of PBSs, MUEs, and PUEs per pico cell\\ \hline
        $M, N_t$  & Numbers of transmit antennas at the MBS and each PBS\\ \hline
        $N_r, N_t$ & Numbers of FD and HD receive antennas and HD\\ \hline
        $\mathbf{h}_{Pk}, \mathbf{H}_{PP}$  & Channel from PBS to $\ms_k$ and self-interference channel at the FD PBS \\ \hline
        $\bar{\mathbf{H}}_{MP}, \bar{\mathbf{h}}_{Mk}$  & Equivalent channels from MBS to PBS and $\ms_k$ \\ \hline
        $\bar{\mathbf{h}}_{MP}, \bar{h}_{Mk}$  & Equivalent channels from MBS to PBS and $\ms_k$ in single-user case \\ \hline
        $\bar{\mathbf{g}}_{MPn}, \bar{g}_{Mkn}$  & Equivalent frequency-domain channels from MBS to PBS and $\ms_k$\\ \hline
        $\mathbf{y}'_P, \mathbf{y}_P$   & Receive signals of FD PBS before and after self-interference cancellation
        \\ \hline
        $\mathbf{z}_x, \mathbf{z}_y$   & Transmitter and receiver distortions from hardware impairments\\ \hline
        $\mathbf{W}_f, L$ & Forwarding precoder at PBS and its order in wideband systems\\ \hline
        $\mathbf{w}_f$ & Vectorization of $\mathbf{W}_f^H$ \\ \hline
        $\mathbf{w}_{d,k}$ & Precoder at PBS for desired signals of $\ms_k$ \\ \hline
        $\bar{\mathbf{W}}_{fn}, \bar{\mathbf{w}}_{dn}$ & Frequency-domain precoders for forwarded ICI and desired signals \\ \hline
        $P_{out}, P_0$ & Total and maximal transmit power of PBS\\ \hline
        $P_{out,n}$ & Transmit power on the $n$-th subcarrier of PBS\\
        % $P_{out}^{fw}$ & Power of PBS allocated for forwarding ICI \\ % \hline
        % $\mathtt{SINR}_{k,FD}, \mathtt{SINR}_{k,HD}$ & SINR of $\ms_k$ in FD and HD modes\\
        % \bottomrule
        \hline
    \end{tabular}
    \end{threeparttable}
\end{table}

In this paper, we consider a HetNet consisting of a MBS and multiple PBSs. We investigate the application of FD technique to mitigate the ICI generated by the MBS to the PUEs. The main contributions are summarized as follows:

\begin{itemize}
  \item We propose a novel ICI cancellation scheme for HetNets, named FD assisted ICI cancellation (fICIC), where the PBSs are supposed to have the FD capability of transmitting and receiving simultaneously. The basic idea of the proposed fICIC is to let the FD PBSs forward the listened interference from the MBS to the PUEs to neutralize the ICI, while at the same time sending the desired signals. Since the ICI is mitigated at PBSs now, the proposed fICIC is transparent to the MBS in the sense that no changes are needed for the transmission of the MBS. Note that a similar usage of the FD technique was presented in~\cite{Zheng2013} for the cooperative cognitive network, where the FD secondary BS forwards the listened primary signals in order to increase the primary spectrum accessing opportunities, which leads to completely different problem and strategy design from ours.
       %Therefore, it is convenient to combine the fICIC with existing eICIC and CoMP-CB techniques that are performed at the MBS if needed.
  \item In narrowband single-user case, where the MBS serves a single MUE and each PBS serves a single PUE, we first find the explicit expressions of the optimal fICIC precoders, which maximize the signal-to-interference-plus-noise ratio (SINR) of the PUE in each pico cell. We then analyze the asymptotical performance of the fICIC in ICI-dominated scenario. The results show that under perfect self-interference cancellation for FD, the fICIC can thoroughly eliminate weak ICI, while when the ICI is very strong or the residual self-interference is very large, the fICIC will reduce to the HD scheme.
  \item In narrowband multi-user case, where the MBS serves multiple MUEs and each PBS serves multiple PUEs, we propose a low-complexity algorithm to optimize the fICIC scheme, aimed at maximizing the downlink sum rate of each pico cell under the fairness constraint over the PUEs.
      %, then we obtain the necessary condition on the number of antennas at the PBSs for fully cancelling the ICI.
      Simulations validate the analytical results, and demonstrate a significant performance gain of the fICIC over the HD scheme as well as evident benefits of combining the fICIC with existing eICIC and CoMP techniques.
  \item We generalize the narrowband fICIC to orthogonal frequency-division multiplexing (OFDM) systems, where the optimization problem for the fICIC precoder design aimed at maximizing sum rate over multiple subcarriers is obtained, which can be solved with a gradient based method. Simulations shows the advantages of the fICIC on mitigating wideband ICI.
\end{itemize}

\textbf{Notations:} $(\cdot)^T$, $(\cdot)^\ast$, and $(\cdot)^H$ denote transpose, complex conjugate, and conjugate
transpose, respectively. $\mathbf{X} \succeq \mathbf{0}$ represents that matrix $\mathbf{X}$ is positive semi-definite, and $\mathbf{X}^{\frac{1}{2}}$ denotes hermitian square root of $\mathbf{X}$. $\mathtt{tr}(\cdot)$ denotes matrix trace, $\mathtt{rank}(\cdot)$ denotes matrix rank, $\mathcal{E}\{\cdot\}$ denotes expectation operator, $\mathtt{vec}(\cdot)$ denotes vectorization operator, $\otimes$ denotes Kronecker product, $\odot$ denotes convolution product, and $\|\cdot\|$ denotes Euclidian norm. $\mathtt{diag}(\mathbf{X})$ denotes the diagonal matrix with the same diagonal elements as $\mathbf{X}$. $\mathbf{I}_N$, $\mathbf{0}_N$ and $\bar{\mathbf{0}}_N$ denote $N\times N$ identity and zero matrices, and $N\times 1$ zero vector, respectively. The major symbols used in the paper are listed in Table~\ref{T:Acronyms}.

\section{System Model} \label{S:system_model}
We first consider the downlink transmission of a narrowband time division duplex (TDD) HetNet, and then generalize the model to wideband systems. Suppose that the HetNet consists of one MBS and $B$ PBSs, where the MBS serves ${K_M}$ single-antenna MUEs and each PBS serves ${K_P}$ single-antenna PUEs. Assume that the MUEs experience negligible interference from PBSs due to the coverage range expansion (CRE) of pico cells, and the pico cells are geographically separated so that each PUE receives much weaker interference from interfering PBSs compared to the interference generated by the MBS, which is treated as noise in the paper. Therefore, we focus on the suppression of the cross-tier ICI generated by the MBS to PUEs, which is commonly recognized as a bottleneck to improve the spectral efficiency in real-world HetNets~\cite{Soret2013}.

\begin{figure*}[!t]
\centering \subfigure[]{\includegraphics[width=0.65\textwidth]{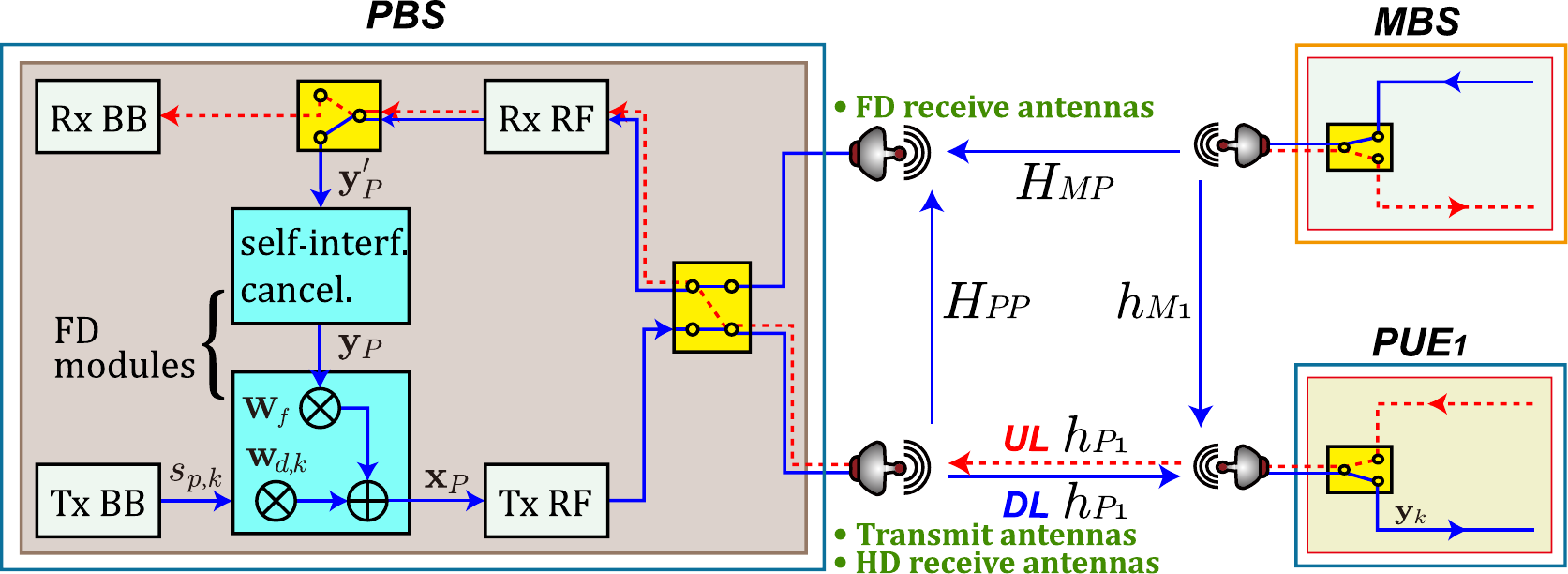}
\label{F:system}}\ \ \  \subfigure[]{\includegraphics[width=0.3\textwidth, height=0.3\textwidth]{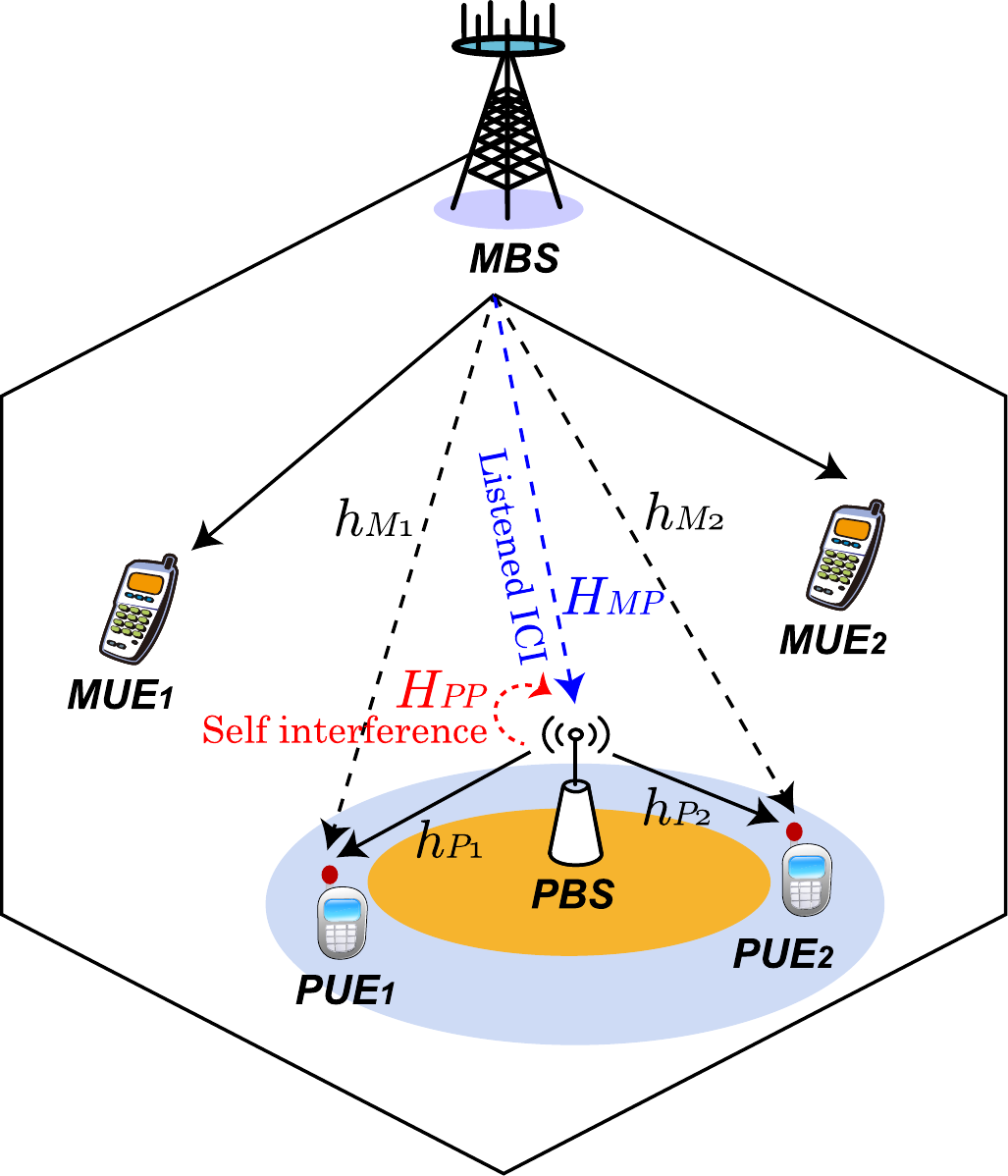}
\label{F:network}} \caption{(a) System model of the TDD HetNet, where the transceiver structure of the FD PBS is illustrated in the left part. (b) Illustration of the considered HetNet layout, where the MBS serves two MUEs and the reference PBS serves two PUEs.}
\end{figure*}

We consider applying FD technique at each PBS in the downlink transmission, with which the PBS can send the desired signals and forward the listened cross-tier ICI simultaneously. The structure of the FD PBS transceiver is illustrated in Fig.~\ref{F:system}, where uplink and downlink data flows are indicated by dashed and solid lines, respectively. The FD PBS is comprised of traditional baseband (BB) and radio frequency (RF) modules in HD transceiver, taking charge of the transmission and reception of desired signals of PUEs, as well as additional FD modules, dedicated for self-interference cancellation and ICI suppression. The antennas at the PBS can be divided into two parts as shown in Fig.~\ref{F:system}, where the ``FD receive antennas'' are only used to receive the ICI from the MBS in the downlink, while the ``transmit antennas'' and the ``HD receive antennas'' share the same antennas in a TDD manner, which are used to send and receive the signals of PUEs in downlink and uplink, respectively. Therefore, the uplink and downlink channel reciprocity holds between the PBS and the PUE.

%\begin{figure}
%\centering
%\includegraphics[width=0.6\textwidth]{Smallcell1} \caption{System model of the TDD HetNet, where the transceiver structure of the FD PBS is illustrated in the left part.}\label{F:system}
%\vspace{-0.6cm}
%\end{figure}

Since the proposed fICIC scheme will not affect the performance of MUEs and other-cell PUEs, in the sequel we only consider a reference PBS and focus on the performance of the PUEs served by the reference PBS. The resulting interference environment is demonstrated in Fig.~\ref{F:network}.
Suppose that the MBS has $M$ transmit antennas, $M \geq {K_M}$, and the FD PBS has $N_t$ transmit antennas, $N_t$ HD receive antennas, and $N_r$ FD receive antennas, $N_t \geq {K_P}$. Let $\mathbf{h}_{Mk}\in\mathbb{C}^{M\times 1}$ and $\mathbf{h}_{Pk}\in\mathbb{C}^{N_t\times 1}$ denote the channels from the MBS and the PBS to the $k$-th PUE (denoted by $\ms_k$), $\mathbf{H}_{MP}\in\mathbb{C}^{M\times N_r}$ denote the channel from the MBS to the PBS, and $\mathbf{H}_{PP}\in\mathbb{C}^{N_t\times N_r}$~denote the self-interference channel of the FD~PBS.

%\begin{figure}[!htb]
%\centering
%\includegraphics[width=0.4\textwidth]{HetNet} \caption{Illustration of the considered HetNet layout, where the MBS serves two MUEs and the reference PBS serves two PUEs.}\label{F:network}
%\vspace{-0.6cm}
%\end{figure}
\subsection{Signals of the FD PBS} \label{S:system_model_1}
In the downlink the FD PBS can transmit and receive signals simultaneously. We use the index~$t$ to denote time instant. To reflect the impact of hardware impairments of transmitter chains on self-interference cancellation, we can express the receive signal at the PBS before self-interference cancellation based on~\cite{Day2012,Zheng2013}~as
\begin{equation} \label{E:receivesignal}
  \bar{\mathbf{y}}_p[t] = \mathbf{H}_{MP}^H \mathbf{W}_M \mathbf{s}_M[t] + \mathbf{H}_{PP}^H \big(\mathbf{x}_p[t] + \mathbf{z}_x[t]\big) + \mathbf{n}_p[t],
\end{equation}
where $\mathbf{W}_M\in\mathbb{C}^{M\times {K_M}}$ is the precoding matrix at the MBS for sending the signals $\mathbf{s}_M\sim\mathcal{CN}(\bar{\mathbf{0}}_{K_M}, \mathbf{I}_{{K_M}})$ to the MUEs, the term $\mathbf{H}_{PP}^H \big(\mathbf{x}_p[t] + \mathbf{z}_x[t]\big)$ is the self-interference, $\mathbf{x}_p\in\mathbb{C}^{N_t\times 1}$ is the transmit signal vector of the FD PBS, $\mathbf{z}_x\sim\mathcal{CN}(\bar{\mathbf{0}}_{N_t}, \mu_x\mathtt{diag}(\boldsymbol{\Phi}_{x}))$ is the transmitter distortion with $\boldsymbol{\Phi}_{x}$ denoting the covariance matrix of $\mathbf{x}_p$, $\mu_x\ll 1$ is a scaling constant, which reflects the combined effects of additive power-amplifier noise, non-linearities in digital-to-analog converter and power amplifier, I/Q imbalance, and oscillator phase noise, and $\mathbf{n}_p\!\sim\!\mathcal{CN}(\bar{\mathbf{0}}_{N_r}, \sigma_n^2\mathbf{I}_{N_r})$ is the additive white Gaussian noise (AWGN), which takes into account both thermal noises and the ICI from other PBSs.

Further considering the hardware impairments of receiver chains based on~\cite{Day2012,Zheng2013}, the distorted receive signal can expressed as
\begin{equation} \label{E:distorted-y}
  \mathbf{y}'_p[t] = \bar{\mathbf{y}}_p[t] + \mathbf{z}_y[t],
\end{equation}
where $\mathbf{z}_y\sim\mathcal{CN}(\bar{\mathbf{0}}_{N_r}, \mu_y\mathtt{diag}(\boldsymbol{\Phi}_y))$ is the additive distortion caused by adaptive gain control noise, non-linearities in analog-to-digital converter and gain control, I/Q imbalance, and oscillator phase noise in receiver chains, $\mu_y\ll 1$ is a scaling constant, and $\boldsymbol{\Phi}_y$ is the covariance matrix of the undistorted receive signal $\bar{\mathbf{y}}_p$, which can be obtained from (\ref{E:receivesignal})~as
\begin{align} \label{E:Phi-y}
  \boldsymbol{\Phi}_y & = \mathcal{E}_{\mathbf{s}_M, \mathbf{x}_{p}, \mathbf{z}_x,\mathbf{n}_p}\{\bar{\mathbf{y}}_p[t]\bar{\mathbf{y}}_p^H[t]\} = \mathbf{H}_{MP}^H \mathbf{W}_M\mathbf{W}_M^H\mathbf{H}_{MP} \nonumber\\
  & + \mathbf{H}_{PP}^H (\boldsymbol{\Phi}_x+\mu_x\mathtt{diag}(\boldsymbol{\Phi}_x)) \mathbf{H}_{PP} + \sigma_n^2\mathbf{I}_{N_r}.
\end{align}

Since the transmit signal $\mathbf{x}_p$ is known at the FD PBS, the self-interference $\mathbf{H}_{PP}^H\mathbf{x}_p$ in (\ref{E:receivesignal}) can be cancelled if the self-interference channel $\mathbf{H}_{PP}$ is estimated. During a dedicated training phase, the orthogonal pilot signals $\sqrt{P_{tr}}\mathbf{C}$ are transmitted for self-interference channel estimation, where $P_{tr}$ is the transmit power of pilot signals, and $\mathbf{C}=[\mathbf{c}_1,\dots,\mathbf{c}_{N_t}]\in\mathbb{C}^{N_t\times N_t}$ is unitary with $\mathbf{C}\mathbf{C}^H = \mathbf{I}_{N_t}$. Then, similar to~(\ref{E:receivesignal}) and (\ref{E:distorted-y}), we can obtain the receive pilot signals with hardware impairments of both transmitter and receiver chains as
\begin{equation} \label{E:Training}
  \mathbf{Y}[t]\!=\! \tilde{\mathbf{Y}}[t] \!+ \!\tilde{\mathbf{Z}}_{y}[t] \!\triangleq\! \mathbf{H}_{PP}^H \big(\sqrt{P_{tr}}\mathbf{C}\!+\!\tilde{\mathbf{Z}}_{x}[t]\big)\! +\! \tilde{\mathbf{N}}[t] \!+\! \tilde{\mathbf{Z}}_{y}[t],
\end{equation}
where $\tilde{\mathbf{Y}}= [\tilde{\mathbf{y}}_{1},\dots,\tilde{\mathbf{y}}_{N_t}] \in \mathbb{C}^{N_r\times N_t}$ denotes the undistorted receive signal, $\tilde{\mathbf{Z}}_y=[\tilde{\mathbf{z}}_{y,1},\dots,\tilde{\mathbf{z}}_{y,N_t}] \in \mathbb{C}^{N_r\times N_t}$ is the additive distortion of receive signals with $\tilde{\mathbf{z}}_{y,i}\sim\mathcal{CN}(\bar{\mathbf{0}}_{N_r}, \mu_y\mathtt{diag}(\tilde{\boldsymbol{\Phi}}_{y,i}))$, $\tilde{\boldsymbol{\Phi}}_{y,i}$ is the covariance matrix of $\tilde{\mathbf{y}}_{i}$, $\tilde{\mathbf{Z}}_x=[\tilde{\mathbf{z}}_{x,1},\dots,\tilde{\mathbf{z}}_{x,N_t}] \in \mathbb{C}^{N_t\times N_t}$ is the transmitter distortion with $\tilde{\mathbf{z}}_{x,i} \sim \mathcal{CN}(\bar{\mathbf{0}}_{N_t}, \mu_xP_{tr}\mathtt{diag}(\mathbf{c}_i\mathbf{c}_i^H))$, and $\tilde{\mathbf{N}} \in \mathbb{C}^{N_r\times N_t}$ is the AWGN whose columns follow the distribution $\mathcal{CN}(\bar{\mathbf{0}}_{N_r}, \sigma_n^2\mathbf{I}_{N_r})$. It can be obtained from (\ref{E:Training}) that $\tilde{\boldsymbol{\Phi}}_{y,i} = P_{tr}\mathbf{H}_{PP}^H \big(\mathbf{c}_i\mathbf{c}_i^H+\mu_x\mathtt{diag}(\mathbf{c}_i\mathbf{c}_i^H)\big)\mathbf{H}_{PP} + \sigma_n^2\mathbf{I}_{N_r}$.

With least-squares channel estimator, we can estimate the self-interference channel as
\begin{align} \label{E:CE}
  \hat{\mathbf{H}}_{PP} = &\frac{1}{\sqrt{P_{tr}}}\mathbf{C}\mathbf{Y}^H[t] = \mathbf{H}_{PP} + \frac{1}{\sqrt{P_{tr}}}\big(\mathbf{C}\tilde{\mathbf{Z}}_x^H[t]\mathbf{H}_{PP} + \nonumber\\
   &\mathbf{C}\tilde{\mathbf{N}}^H[t] + \mathbf{C}\tilde{\mathbf{Z}}_y^H[t]\big) \triangleq \mathbf{H}_{PP} + \mathbf{E}_{PP}[t],
\end{align}
where $\mathbf{E}_{PP} = [\mathbf{e}_{PP,1},\dots,\mathbf{e}_{PP,N_t}]^H \in \mathbb{C}^{N_t\times N_r}$ denotes the channel estimation errors.

Denoting $\mathbf{c}_i = [c_{i1}, \dots, c_{iN_t}]^T$, we can express $\mathbf{e}_{PP,i}$ as
\begin{equation}
  \mathbf{e}_{PP,i} = \frac{1}{\sqrt{P_{tr}}} \Big(\mathbf{H}_{PP}^H \sum_{j=1}^{N_t}{c}_{ij}\tilde{\mathbf{z}}_{x,j} + \tilde{\mathbf{N}}\mathbf{c}_i + \sum_{j=1}^{N_t} c_{ij}\tilde{\mathbf{z}}_{y,j}\Big),
\end{equation}
from which we can obtain that $\mathbf{e}_{PP,i}$ follows the distribution $\mathcal{CN}(\bar{\mathbf{0}}_{N_r}, \tilde{\boldsymbol{\Phi}}_{e,i})$ with
\begin{align} \label{E:Epp-cov}
 & \tilde{\boldsymbol{\Phi}}_{e,i}
  %&=\frac{1}{P_{tr}}\Big(\mathbf{H}_{PP}^H \sum_{j=1}^{N_t}|c_{ij}|^2\mu_xP_{tr}\mathtt{diag}(\mathbf{c}_i\mathbf{c}_i^H) \mathbf{H}_{PP} + \sigma_n^2\mathbf{I}_{N_r} \nonumber\\
  %&+ \sum_{j=1}^{N_t} |c_{ij}|^2 \mu_yP_{tr}\mathtt{diag}\big(\mathbf{H}_{PP}^H \big(\mathbf{c}_i\mathbf{c}_i^H+\mu_x\mathtt{diag}(\mathbf{c}_i\mathbf{c}_i^H)\big)\mathbf{H}_{PP}\big) + |c_{ij}|^2\mu_y\sigma_n^2\mathbf{I}\Big)\nonumber\\
 \!=\! \mathbf{H}_{PP}^H \sum_{j=1}^{N_t}|c_{ij}|^2\mu_x\mathtt{diag}(\mathbf{c}_j\mathbf{c}_j^H) \mathbf{H}_{PP} \! +\!  \frac{(1\! +\! \mu_y)\sigma_n^2}{P_{tr}}\mathbf{I}_{N_r}\! +\! \nonumber\\
  & \sum_{j=1}^{N_t} |c_{ij}|^2 \mu_y\mathtt{diag}\big(\mathbf{H}_{PP}^H \big(\mathbf{c}_j\mathbf{c}_j^H\! +\! \mu_x\mathtt{diag}(\mathbf{c}_j\mathbf{c}_j^H)\big)\mathbf{H}_{PP}\big).
\end{align}

%\begin{align} \label{E:Epp-cov}
%  \tilde{\boldsymbol{\Phi}}_{e,i}
%  = &\frac{1}{P_0}\Big(\mathbf{H}_{PP}^H \mu_x\mathtt{diag}(\mathbf{c}_i\mathbf{c}_i^H) \mathbf{H}_{PP} + (1+\mu_y) \sigma_n^2\mathbf{I}_{N_r} \nonumber\\
%  &\qquad\qquad+ \mu_y\mathtt{diag}\big(\mathbf{H}_{PP}^H \big(\mathbf{c}_i\mathbf{c}_i^H+\mu_x\mathtt{diag}(\mathbf{c}_i\mathbf{c}_i^H)\big)\mathbf{H}_{PP}\big)\Big).
%  %= &\mathbf{H}_{PP}^H \big(\mathbf{c}_i\mathbf{c}_i^H+2\mu_x\mathtt{diag}(\mathbf{c}_i\mathbf{c}_i^H)\big)\mathbf{H}_{PP} + 2\sigma_n^2\mathbf{I}_{N_r}.
%\end{align}

With $\hat{\mathbf{H}}_{PP}$, the receive signal at the PBS after self-interference cancellation is
\begin{align} \label{E:yp}
  \mathbf{y}_p[t] = & \mathbf{y}_p'[t] - \hat{\mathbf{H}}_{PP}^H[t] \mathbf{x}_p[t] \triangleq
  \bar{\mathbf{H}}_{MP}^H \mathbf{s}_M[t] - \mathbf{E}_{PP}^H[t]\mathbf{x}_p[t] \nonumber\\
  & + \mathbf{H}_{PP}^H \mathbf{z}_x[t] + \mathbf{n}_p[t] + \mathbf{z}_y[t],
\end{align}
where $\bar{\mathbf{H}}_{MP} \triangleq \mathbf{W}_M^H\mathbf{H}_{MP}$ is the equivalent channel from the MBS to the FD PBS.

The FD PBS then transmits the desired signals of the PUEs together with the self-interference cancelled receive signals $\mathbf{y}_p$. The combined transmit signal of the PBS can be expressed as
\begin{equation}\label{E:x2}
  \mathbf{x}_p[t] = \mathbf{W}_f \mathbf{y}_p[t-\tau] + \sum_{k=1}^{K_P} \mathbf{w}_{d,k} s_{p,k}[t],
\end{equation}
where $\tau$ is the processing delay introduced by the FD modules, $\mathbf{W}_f\in\mathbb{C}^{N_t\times N_r}$ is the precoding matrix for the forwarded signals, $\mathbf{w}_{d,k}\in\mathbb{C}^{N_t\times 1}$ is the precoding vector for the desired signal, $s_{p,k}$, of $\ms_k$, and $s_{p,k}\sim\mathcal{CN}(0,1)$.

In (\ref{E:x2}), the receive signal $\mathbf{y}_p$ by $N_r$ additional FD receive antennas of the PBS are forwarded via $\mathbf{W}_f$. The forwarded signal will occupy a part of transmit power of the PBS, which leads to the reduction of the transmit power for desired signals. As will be clear later, however, the forwarded signal can be used to  mitigate the cross-tier ICI at PUEs efficiently with the optimized $\mathbf{W}_f$, and hence result in the improvement of PUEs' data rate.

With (\ref{E:yp}) and (\ref{E:x2}) we can calculate the transmit power of the FD PBS as
\begin{equation} \label{E:Pout-new}
    P_{out} \!=\! \mathtt{tr}(\boldsymbol{\Phi}_x) \!=\! \mathtt{tr}\big(\mathcal{E}_{\mathbf{s}_M, s_{p,k}, \mathbf{n}_p, \mathbf{z}_x,\mathbf{z}_y, \mathbf{E}_{PP},\mathbf{H}_{PP}}\{\mathbf{x}_p[t]\mathbf{x}_p^H[t]\}\big),
\end{equation}
where the expectations are taken over data $\mathbf{s}_M$ and $s_{p,k}$, noises $\mathbf{n}_p$, transmitter and receiver distortions $\mathbf{z}_x$ and $\mathbf{z}_y$, channel estimation errors $\mathbf{E}_{PP}$, and self-interference channel $\mathbf{H}_{PP}$, respectively.\footnote{In order to highlight the benefits of the fICIC scheme via forwarding the listened ICI, we restrict ourselves to the case where the precoders $\mathbf{W}_f$ and $\{\mathbf{w}_{d,k}\}$ are designed only for transmitting the listened ICI and desired signals but not for spatial-domain self-interference cancellation (i.e., design $\mathbf{W}_f$ and $\{\mathbf{w}_{d,k}\}$ to pre-null self-interference). Towards this end, we take the expectation over $\mathbf{H}_{PP}$ in (\ref{E:Pout-new}), such that the precoders are independent of the instantaneous self-interference channel~$\mathbf{H}_{PP}$. }

Assume that $\mathbf{H}_{PP}$ follows Rayleigh distribution, i.e., $\mathtt{vec}(\mathbf{H}_{PP})\sim\mathcal{CN}(\bar{\mathbf{0}}_{N_rN_t}, \bar{\alpha}_{PP}\mathbf{I}_{N_rN_t})$ with $\bar{\alpha}_{PP}$ denoting the average channel gain, which is reasonable because the transmit antennas and the FD receive antennas can be well isolated in the considered scenario as will be detailed in Section~\ref{S:SIC}. Then, we show in Appendix~\ref{A:Pout} that the transmit power $P_{out}$ can be expressed~as
\begin{align}\label{E:x22}
  P_{out} \approx & \mathtt{tr}\left(\mathbf{W}_f\bar{\mathbf{H}}_{MP}^H\bar{\mathbf{H}}_{MP}\mathbf{W}_f^H\right) +
  \sum_{k=1}^{K_P} \|\mathbf{w}_{d,k}\|^2\nonumber\\
  & + \big(\sigma_n^2 + \sigma_e^2P_{out} \big) \mathtt{tr}(\mathbf{W}_f\mathbf{W}_f^H),
\end{align}
where the approximation follows from $\mu_x\ll 1$ and $\mu_y\ll 1$ as in~\cite{Day2012}, and $\sigma^2_e = \frac{\sigma_n^2}{P_{tr}}+2\bar{\alpha}_{PP}(\mu_x+\mu_y)$ reflects the residual self-interference, in which the term $\frac{\sigma_n^2}{P_{tr}}$ comes from imperfect channel estimation and the term $2\bar{\alpha}_{PP}(\mu_x+\mu_y)$ comes from hardware impairments.

From the equation with respect to $P_{out}$ given in (\ref{E:x22}), we can obtain the transmit power of the FD PBS as
\begin{align} \label{E:Pout}
  & P_{out} =\\
  & \frac{\mathtt{tr}\!\big(\bar{\mathbf{H}}_{MP}\mathbf{W}_f^H\mathbf{W}_f\bar{\mathbf{H}}_{MP}^H\big) \!+\! \sigma_n^2\mathtt{tr}\!\big(\mathbf{W}_f^H\mathbf{W}_f\big) \!+\! \sum_{k=1}^{K_P}\! \|\mathbf{w}_{d,k}\|^2}{1 \!-\! \sigma_e^2\mathtt{tr}\big(\mathbf{W}_f^H\mathbf{W}_f\big)},\nonumber
\end{align}
which shows that the power of the precoding matrix $\mathbf{W}_f$ for ICI forwarding must be limited~by
\begin{equation} \label{E:oscillations}
  \mathtt{tr}\left(\mathbf{W}_f^H\mathbf{W}_f\right) < \frac{1}{\sigma_e^2}.
\end{equation}
Otherwise, amplifier self-oscillations will occur at the FD PBS~\cite{Riihonen2009}.

Let $P_0$ denote the maximal transmit power of the PBS. Then the transmit power $P_{out}$ needs to satisfy $P_{out} \leq P_0$, which can be rewritten based on (\ref{E:Pout}) as
\begin{align} \label{E:PBPC}
  & \mathtt{tr}\left(\bar{\mathbf{H}}_{MP}\mathbf{W}_f^H\mathbf{W}_f\bar{\mathbf{H}}_{MP}^H\right) + (P_0\sigma_e^2+ \sigma_n^2)\mathtt{tr}\left(\mathbf{W}_f^H\mathbf{W}_f\right) \nonumber\\
  & \qquad\qquad\qquad\qquad\qquad\quad\quad + \sum_{k=1}^{K_P} \|\mathbf{w}_{d,k}\|^2 \leq P_0.
\end{align}

One can observe that the self-oscillation constraint (\ref{E:oscillations}) is implicitly reflected in the maximal power constraint (\ref{E:PBPC}). Therefore, in the following only the constraint in (\ref{E:PBPC}) is considered.

\subsection{Signals of $\ms_k$}
The receive signal of $\ms_k$ can be expressed as
\begin{equation}\label{E:yk}
  y_k[t] = \mathbf{h}_{Pk}^H\mathbf{x}_p[t] + \mathbf{h}_{Mk}^H\mathbf{W}_M\mathbf{s}_M[t] + n_k[t],
\end{equation}
where the impact of hardware impairments at the HD $\ms$ is ignored as commonly considered for HD transmission in the literature, and $n_k[t]\sim\mathcal{CN}(0, \sigma_n^2)$ is the AWGN at $\ms_k$, which includes the received ICI from interfering PBSs and thermal noises.

By noting the equivalence between time delay and phase shift in narrowband systems, we have $\mathbf{s}_M[t-\tau] = \mathbf{s}_M[t]e^{-j\phi}$, where $\phi = 2\pi f_c\tau$ with $f_c$ denoting the carrier frequency.\footnote{One can further incorporate the propagation delay difference experienced by the forwarded ICI and the direct ICI into $\tau$, which is not considered here because the PUEs are close to the PBS leading to negligible additional propagation delay.} Then based on (\ref{E:yp}) and (\ref{E:x2}), we can rewrite (\ref{E:yk}) as
\begin{align}\label{E:yk2}
    % y_k[t] &= \mathbf{h}_{Pk}^H \sum_{k=1}^{K_P} \mathbf{w}_{d,k} s_{p,k}[t] + \mathbf{h}_{Pk}^H \mathbf{w}_f ( \mathbf{h}_{MP}^H \mathbf{W}_M \mathbf{s}_M[t-\tau] + \mathbf{e}_{PP}^H\mathbf{x}_p[t-\tau] + n_p[t-\tau] ) +  \mathbf{h}_{Mk}^H\mathbf{W}_M\mathbf{s}_M[t] + n_k[t].\\
 & y_k[t] = \mathbf{h}_{Pk}^H  \mathbf{w}_{d,k} s_{p,k}[t] +  \underbrace{\sum_{j=1, j\neq k}^{K_P} \mathbf{h}_{Pk}^H \mathbf{w}_{d,j} s_{p,j}[t]}_{\text{Intra-cell interference}} \nonumber\\
& + \mathbf{h}_{Pk}^H \mathbf{W}_f \big(-\mathbf{E}_{PP}^H[t-\tau]\mathbf{x}_p[t-\tau] + \mathbf{H}_{PP}^H \mathbf{z}_x[t-\tau] +\nonumber\\
  &\quad\underbrace{\qquad\qquad\qquad\qquad\qquad\qquad\quad \mathbf{n}_p[t-\tau] + \mathbf{z}_y[t-\tau]\big)}_{\text{Forwarded residual self-interference and noises}} \nonumber\\
   & + \underbrace{\left( \bar{\mathbf{h}}_{Mk}^H + \mathbf{h}_{Pk}^H \mathbf{W}_f \bar{\mathbf{H}}_{MP}^H e^{-j\phi} \right) \mathbf{s}_M[t]}_{\text{Cross-tier ICI}}
  +  n_k[t],
\end{align}
where $\bar{\mathbf{h}}_{Mk} \triangleq \mathbf{W}_M^H\mathbf{h}_{Mk}$ is the equivalent channel from the MBS to $\ms_k$.

To achieve coherent combination between the ICI forwarded by the PBS and that directly received at the PUEs as shown in (\ref{E:yk2}), the PBS needs to implement sample-by-sample ICI forwarding in time domain to reduce the processing delay as considered in~\cite{Riihonen2008a, Riihonen2008} for relay systems, where the time-domain FD amplify-and-forward relay was employed to provide co-phasing combining gain at the destination node. This is different from frequency domain forwarding for FD relay systems, e.g., in\cite{Riihonen2009} by the same authors as~\cite{Riihonen2008a, Riihonen2008}, where the processing delay will be in symbol-level because a complete symbol needs to be received and demodulated before forwarding in the frequency domain.

Similar to the derivations in Appendix~\ref{A:Pout}, we can compute the signal and interference power from (\ref{E:yk2}) by taking the expectation with respect to $\mathbf{s}_M$, $s_{p,k}$, $\mathbf{n}_p$, $\mathbf{z}_x$, $\mathbf{z}_y$, $\mathbf{E}_{PP}$, and $\mathbf{H}_{PP}$, and finally obtain the SINR of $\ms_k$ as
\begin{align}\label{E:SINR3}
  &\mathtt{SINR}_{k,FD}= \nonumber\\
  & |\mathbf{h}_{Pk}^H\mathbf{w}_{d,k}|^2 \!/\! \big(\sum_{j\neq k}\! \|\mathbf{h}_{Pk}^H \mathbf{w}_{d,j}\|^2 \!+\! \|\bar{\mathbf{h}}_{Mk}^H \!+\! \mathbf{h}_{Pk}^H \mathbf{W}_f \bar{\mathbf{H}}_{MP}^H e^{-j\phi}\!\|^2  \nonumber\\
  &+ \|\mathbf{h}_{Pk}^H \mathbf{W}_f\|^2(P_{out}\sigma_e^2+\sigma_n^2) + \sigma_n^2\big),
\end{align}
where $P_{out}$ is the transmit power of the PBS, which is a function of the precoders $\mathbf{W}_f$ and $\mathbf{w}_{d,k}$ as given in~(\ref{E:Pout}).

%\subsection{Mechanism of fICIC}
%The fICIC scheme has two mechanisms to suppress the cross-tier ICI, namely \emph{ICI-weaken} and \emph{ICI-enhance}. With the ICI-weaken mechanism, the forwarded signals are designed to reduce the ICI, then the PUEs simply treat the ICI as noises when decoding. With the ICI-enhance mechanism, the forwarded signals are designed to enhance the ICI so that non-linear successive interference cancellation (SIC) can be used at the PUEs. In the paper, both the mechanisms are studied for single-user and multi-user case, respectively.

\section{Narrowband Single-user Case}
In this section, we investigate the single-user case, i.e., ${K_M}={K_P}=1$, to gain insight into the ICI mitigation mechanism of the fICIC scheme.

In this case, the equivalent channel $\bar{\mathbf{H}}_{MP}$ is a row vector and $\bar{\mathbf{h}}_{Mk}$ is a scalar, which are denoted by $\bar{\mathbf{h}}_{MP}^H \in \mathbb{C}^{1\times N_r}$ and $\bar{h}_{Mk}\in \mathbb{C}^{1\times 1}$ for clarity, respectively. Moreover, noting that the intra-cell interference does not exist now, the SINR can be simplified as
\begin{align}\label{E:SINR3-1}
  &\mathtt{SINR}_{k,FD} =\\
  &\frac{|\mathbf{h}_{Pk}^H\mathbf{w}_{d,k}|^2}{ |\bar{h}_{Mk}^\ast \!+\! \mathbf{h}_{Pk}^H \mathbf{W}_f \bar{\mathbf{h}}_{MP} e^{-j\phi}|^2  \!+\! \|\mathbf{h}_{Pk}^H \mathbf{W}_f\|^2\!(P_{out}\sigma_e^2\!+\!\sigma_n^2) \!+\! \sigma_n^2}, \nonumber
\end{align}
where $P_{out}$ given in (\ref{E:Pout}) can be rewritten as
\begin{equation} \label{E:pout-2}
  P_{out} = \frac{\|\mathbf{W}_f\bar{\mathbf{h}}_{MP}\|^2  + \sigma_n^2\mathtt{tr}\left(\mathbf{W}_f^H\mathbf{W}_f\right) +  \|\mathbf{w}_{d,k}\|^2}{1 - \sigma_e^2\mathtt{tr}\left(\mathbf{W}_f^H\mathbf{W}_f\right)}.
\end{equation}

When the precoder for forwarding the listened ICI, $\mathbf{W}_f$, in (\ref{E:SINR3-1}) is selected as zero, the FD PBS reduces to a HD PBS. It is not hard to obtain the optimal precoder for transmitting the desired signal in HD case that maximizes the SINR of $\ms_k$ as $\mathbf{w}_{d,k} = \frac{\sqrt{P_0}\mathbf{h}_{Pk}}{\|\mathbf{h}_{Pk}\|}$, which is referred to as ``the HD scheme'' in the sequel. The maximum SINR achieved by the HD scheme can be obtained from (\ref{E:SINR3-1}) as
\begin{equation}\label{E:SINR3-HD}
  \mathtt{SINR}_{k,HD}^\star = \frac{P_0\|\mathbf{h}_{Pk}\|^2}{|\bar{h}_{Mk}|^2 + \sigma_n^2}.
\end{equation}

Compared to the HD scheme where the ICI power is $|\bar{h}_{Mk}|^2$ as shown in (\ref{E:SINR3-HD}), the FD scheme turns the ICI power into $|\bar{h}_{Mk}^\ast + \mathbf{h}_{Pk}^H \mathbf{W}_f \bar{\mathbf{h}}_{MP} e^{-j\phi}|^2$ as shown in (\ref{E:SINR3-1}), which can be reduced by optimizing the precoders $\mathbf{W}_f$ and $\mathbf{w}_{d,k}$. The optimization problem, aimed at maximizing the SINR of $\ms_k$ subject to the maximal transmit power constraint, can be formulated as
\begin{subequations} \label{E:problem}
    \begin{align}
        \underset{\mathbf{W}_f, \mathbf{w}_{d,k}}{\max}\ & \mathtt{SINR}_{k,FD}  \label{E:Objective}\\
         s.\ t.\ &  \|\mathbf{W}_f\bar{\mathbf{h}}_{MP}\|^2 + (P_0\sigma_e^2 + \sigma_n^2)\mathtt{tr}\left(\mathbf{W}_f^H\mathbf{W}_f\right) \nonumber\\
         &\qquad\qquad\qquad\qquad\qquad+ \|\mathbf{w}_{d,k}\|^2 \leq P_0, \label{E:Constraint}
    \end{align}
\end{subequations}
where the power constraint (\ref{E:Constraint}) comes from (\ref{E:PBPC}).

\subsection{Optimal fICIC Scheme}
In this subsection, we strive to find the optimal fICIC scheme with explicit expressions for $\mathbf{W}_f$ and $\mathbf{w}_{d,k}$. Given that it is difficult to directly solve problem (\ref{E:problem}) due to the complex expression of $\mathtt{SINR}_{k,FD}$ in (\ref{E:SINR3-1}), we start with investigating the properties of the optimal solutions to problem~(\ref{E:problem}) in the following.

First, by substituting the expression of $P_{out}$ given in (\ref{E:pout-2}) into (\ref{E:SINR3-1}), we can find that $\mathtt{SINR}_{k,FD}$, i.e., the objective function of problem (\ref{E:problem}), is an increasing function of $\|\mathbf{w}_{d,k}\|^2$. Therefore, for any given $\mathbf{W}_{f}$ and the direction of $\mathbf{w}_{d,k}$, $\frac{\mathbf{w}_{d,k}}{\|\mathbf{w}_{d,k}\|}$, we can always improve the SINR by increasing $\|\mathbf{w}_{d,k}\|^2$ until the PBS transmits with its maximum power. This means that the global optimal solution to problem (\ref{E:problem}) is obtained when the power constraint in (\ref{E:Constraint}) holds with equality.

Based on this result, we can obtain from (\ref{E:pout-2}) that $P_{out} = P_0$, with which $\mathtt{SINR}_{k,FD}$ is simplified~as
\begin{align}\label{E:SINR3-2}
  &\mathtt{SINR}_{k,FD} = \\
  &\frac{|\mathbf{h}_{Pk}^H\mathbf{w}_{d,k}|^2}{ |\bar{h}_{Mk}^\ast \!+\! \mathbf{h}_{Pk}^H \mathbf{W}_f \bar{\mathbf{h}}_{MP} e^{-j\phi}|^2  \!+\! \|\mathbf{h}_{Pk}^H \mathbf{W}_f\|^2\!(P_0\sigma_e^2\!+\!\sigma_n^2) \!+\! \sigma_n^2}. \nonumber
\end{align}

Second, because the direction of $\mathbf{w}_{d,k}$ affects only the numerator of $\mathtt{SINR}_{k,FD}$, we can obtain that the optimal $\mathbf{w}_{d,k}^\star$ satisfies
\begin{equation} \label{E:w2_direction}
  \frac{\mathbf{w}_{d,k}^\star}{\|\mathbf{w}_{d,k}^\star\|} = \frac{\mathbf{h}_{Pk}}{\|\mathbf{h}_{Pk}\|}.
\end{equation}

Third, we can show that the optimal $\mathbf{W}_{f}^\star$ has the following property.

\emph{\textbf{Lemma 1:} The optimal $\mathbf{W}_{f}^\star$ is of rank 1, which can be decomposed as}
\begin{equation}\label{E:w1}
  \mathbf{W}_{f}^\star = - \bar{h}_{Mk}^\ast e^{j\phi}\cdot \beta \cdot\mathbf{h}_{Pk}\bar{\mathbf{h}}_{MP}^H,
\end{equation}
where $\beta$ is a positive scalar.
\begin{proof}
  See Appendix \ref{S:prooflemma2}.
\end{proof}

We can see from the expression of the optimal $\mathbf{W}_{f}^\star$ that with the fICIC the listened ICI is first enhanced by the maximal ratio combining with $\bar{\mathbf{h}}_{MP}^H$, and then forwarded by the maximal ratio transmission with $\mathbf{h}_{Pk}$.

Based on (\ref{E:w2_direction}) and (\ref{E:w1}), we can express the SINR with $\beta$ and $\|\mathbf{w}_{d,k}\|$. Further considering that the optimal solution is obtained when the power constraint in (\ref{E:Constraint}) holds with equality, we can replace $\|\mathbf{w}_{d,k}\|$ with $\beta$, and finally convert problem (\ref{E:problem}) into the following optimization problem
\begin{subequations} \label{E:problem1}
    \begin{align}
        \underset{\beta}{\max}\ & f_0(\beta)  \label{E:Objective1}\\
         s.\ t.\ &  \left(  \|\bar{\mathbf{h}}_{MP}\|^2 + \sigma_I^2+\sigma_n^2 \right)|\bar{h}_{Mk}|^2\|\mathbf{h}_{Pk}\|^2 \|\bar{\mathbf{h}}_{MP}\|^2 \beta^2 \leq P_0 \label{E:Constraint1}\\
              \  &\beta \geq 0,
    \end{align}
\end{subequations}
where $f_0(\beta) \triangleq \Big(\|\mathbf{h}_{Pk}^H\|^2\big(P_0 - \big(\|\bar{\mathbf{h}}_{MP}\|^2 + \sigma_I^2+\sigma_n^2 \big) |\bar{h}_{Mk}|^2\|\mathbf{h}_{Pk}\|^2 \|\bar{\mathbf{h}}_{MP}\|^2 \beta^2 \big)\Big) / \Big( \big(|\bar{h}_{Mk}| - \beta|\bar{h}_{Mk}|\|\mathbf{h}_{Pk}\|^2 \|\bar{\mathbf{h}}_{MP}\|^2\big)^2  + \beta^2|\bar{h}_{Mk}|^2\|\mathbf{h}_{Pk}\|^4 \|\bar{\mathbf{h}}_{MP}\|^2(\sigma_I^2+\sigma_n^2) + \sigma_n^2\Big)$, and
$\sigma_I^2 \triangleq P_0\sigma_e^2$ denotes the average power of residual self-interference.

\emph{\textbf{Proposition 1:} The maximum of the objective function in (\ref{E:Objective1}), i.e. the maximal SINR of $\ms_k$, can be expressed as}
\begin{equation} \label{E:maxSINR}
  \mathtt{SINR}_{k,FD}^\star = \frac{1}{\frac{1}{B\beta^\star}-1},
\end{equation}
\emph{with $\beta^\star = \frac{2\left(A+D-\sqrt{(A+D)^2 - \frac{AC^2}{B}}\right)}{C^2}$ representing the optimal solution of $\beta$,} % with the expression}
%\begin{equation}\label{E:w1-norm}
%  \beta^\star = \frac{2\left(A+D-\sqrt{(A+D)^2 - \frac{AC^2}{B}}\right)}{C^2},
%\end{equation}
\emph{where $A = P_0\|\mathbf{h}_{Pk}\|^2$, $B = \|\mathbf{h}_{Pk}\|^2 (\|\bar{\mathbf{h}}_{MP}\|^2 + \sigma_I^2 + \sigma_n^2)$, $C = 2|\bar{h}_{Mk}|\|\mathbf{h}_{Pk}\|\|\bar{\mathbf{h}}_{MP}\|$, and $D = |\bar{h}_{Mk}|^2+\sigma_n^2$. }
\begin{proof}
  See Appendix \ref{S:prooftheorem1}.
\end{proof}

With the optimal $\beta^\star$, we can directly obtain the optimal $\mathbf{W}_f^\star$ from (\ref{E:w1}). To compute the optimal $\mathbf{w}_{d,k}^\star$, recalling that constraint (\ref{E:Constraint}) holds with equality for the optimal solutions, we can obtain the norm of the optimal $\mathbf{w}_{d,k}^\star$ as
\begin{equation}\label{E:w2_norm}
  \|\mathbf{w}_{d,k}^\star\| = \sqrt{P_0 - \frac{1}{4}\left(\|\bar{\mathbf{h}}_{MP}\|^2 + \sigma_I^2 + \sigma_n^2\right)C^2\beta^{\star 2}}.
\end{equation}
Then by substituting (\ref{E:w2_norm}) into (\ref{E:w2_direction}), we can obtain $\mathbf{w}_{d,k}^\star$.

\subsection{Asymptotic Performance Analysis} \label{S:asymptotic}
To understand the behavior of the fICIC scheme, we next consider an ICI-dominated scenario, where the noise power $\sigma_n^2$ approaches to~zero.

\subsubsection{Perfect Self-interference Cancellation}
If the self-interference can be perfectly eliminated, i.e., $\sigma_I^2 = 0$, from the definitions after (\ref{E:maxSINR}) we have $B \doteq \|\mathbf{h}_{Pk}\|^2\|\bar{\mathbf{h}}_{MP}\|^2$ and $D \doteq |\bar{h}_{Mk}|^2$, where $\doteq$ denotes asymptotic equality. Then based on Proposition 1, after some manipulations we can obtain the optimal $\beta^\star$ as
\begin{align} \label{E:w1-app}
  &\beta^\star\doteq \frac{P_0\|\mathbf{h}_{Pk}\|^2 + |\bar{h}_{Mk}|^2 - \left|P_0\|\mathbf{h}_{Pk}\|^2 - |\bar{h}_{Mk}|^2\right|}{2\|\bar{\mathbf{h}}_{MP}\|^2|\bar{h}_{Mk}|^2\|\mathbf{h}_{Pk}\|^2}\\
  & = \frac{\min(P_0\|\mathbf{h}_{Pk}\|^2, |\bar{h}_{Mk}|^2)}{\|\bar{\mathbf{h}}_{MP}\|^2|\bar{h}_{Mk}|^2\|\mathbf{h}_{Pk}\|^2} \triangleq \frac{\eta}{\|\bar{\mathbf{h}}_{MP}\|^2|\bar{h}_{Mk}|^2\|\mathbf{h}_{Pk}\|^2}.\nonumber
\end{align}

%By substituting (\ref{E:w1-app}) into (\ref{E:maxSINR}), we can obtain the maximal SINR of $\ms_3$ as
%\begin{equation} \label{E:maxSINR-app}
%  \mathtt{SINR}_{3,FD}^\star \approx \frac{1}{\frac{P_1|h_{Mk}|^2}{\min(P_0|h_{Pk}|^2, P_1|h_{Mk}|^2)}-1}.
%\end{equation}

By substituting (\ref{E:w1-app}) into (\ref{E:Objective1}), the maximal SINR of $\ms_k$ becomes
\begin{equation}\label{E:maxSINR-app}
  \mathtt{SINR}_{k,FD}^\star \doteq \frac{P_0|\bar{h}_{Mk}|^2\|\mathbf{h}_{Pk}\|^2 \!-\! \eta^2}{(\eta \!-\! |\bar{h}_{Mk}|^2)^2 \!+\! \left( \frac{\eta^2}{\|\bar{\mathbf{h}}_{MP}\|^2} \!+\! |\bar{h}_{Mk}|^2\right)\sigma_n^2},
\end{equation}
where $\eta = \min(P_0\|\mathbf{h}_{Pk}\|^2, |\bar{h}_{Mk}|^2)$ as defined in (\ref{E:w1-app}), which depends on the strengths of the desired signal, $P_0\|\mathbf{h}_{Pk}|^2$, and the ICI, $|\bar{h}_{Mk}|^2$. In the following two cases are discussed.

\begin{itemize}
  \item \textbf{Case 1: }$|\bar{h}_{Mk}|^2 < P_0\|\mathbf{h}_{Pk}\|^2$

    This is a case where the ICI is weaker than the desired signal. Then, we have $\eta = |\bar{h}_{Mk}|^2$ and the maximal SINR is
    \begin{equation} \label{E:maxSINR-largeS}
      \overline{\mathtt{SINR}}_{k,FD}^\star \doteq \frac{P_0\|\mathbf{h}_{Pk}\|^2 - |\bar{h}_{Mk}|^2}{\left(\frac{|\bar{h}_{Mk}|^2}{\|\bar{\mathbf{h}}_{MP}\|^2} + 1\right)\sigma_n^2}.
    \end{equation}
    It implies that the FD PBS can thoroughly eliminate the ICI generated by the MBS by properly designing the forwarding and transmitting precoders.

    When compared with the HD scheme, the performance gain of the fICIC can be obtained~as
    \begin{equation} \label{E:FD-HD-1}
      \frac{\overline{\mathtt{SINR}}_{k,FD}^\star}{\mathtt{SINR}_{k,HD}^\star} \doteq \frac{ 1 - \frac{|\bar{h}_{Mk}|^2}{P_0\|\mathbf{h}_{Pk}\|^2}}{\left(\frac{1}{\|\bar{\mathbf{h}}_{MP}\|^2} + \frac{1}{|\bar{h}_{Mk}|^2} \right)\sigma_n^2}.
    \end{equation}

  \item \textbf{Case 2: }$|\bar{h}_{Mk}|^2 \geq P_0\|\mathbf{h}_{Pk}\|^2$

    In this case where the ICI is stronger, we have $\eta = P_0\|\mathbf{h}_{Pk}\|^2$ and the maximal SINR is
    \begin{equation} \label{E:maxSINR-largeI-1}
      \widehat{\mathtt{SINR}}_{k,FD}^\star \doteq \frac{P_0\|\mathbf{h}_{Pk}\|^2}{|\bar{h}_{Mk}|^2\!- \!P_0\|\mathbf{h}_{Pk}\|^2 \!+ \! \frac{\frac{P_0^2\|\mathbf{h}_{Pk}\|^4}{\|\bar{\mathbf{h}}_{MP}\|^2} \!+\! |\bar{h}_{Mk}|^2}{|\bar{h}_{Mk}|^2\!-\!P_0\|\mathbf{h}_{Pk}\|^2}\sigma_n^2}.
    \end{equation}
    For very strong interference, i.e., $|\bar{h}_{Mk}|^2 \gg P_0\|\mathbf{h}_{Pk}\|^2$, $\widehat{\mathtt{SINR}}_{k,FD}^\star$ can be approximated as
    \begin{equation} \label{E:maxSINR-largeI}
      \widehat{\mathtt{SINR}}_{k,FD}^\star \approx \frac{P_0\|\mathbf{h}_{Pk}\|^2}{|\bar{h}_{Mk}|^2} \doteq \mathtt{SINR}_{k,HD}^\star.
    \end{equation}
\end{itemize}

From the above analysis, we can obtain the following observations.
\begin{itemize}
\item \emph{\textbf{Impact of $\bar{\mathbf{h}}_{MP}$}:} It is shown from (\ref{E:maxSINR-largeS}) and (\ref{E:maxSINR-largeI-1}) that the SINR of $\ms_k$ increases with the channel gain between MBS and PBS $\|\bar{\mathbf{h}}_{MP}\|$. This is because given the power of the PBS allocated for forwarding the listened ICI, denoted as
    \begin{equation} \label{E:Powerforward}
      P_{out}^{fw} \triangleq \|\mathbf{W}_f\bar{\mathbf{h}}_{MP}\|^2 + (\sigma_I^2 + \sigma_n^2)\mathtt{tr}\left(\mathbf{W}_f^H\mathbf{W}_f\right),
    \end{equation}
    a large $\|\bar{\mathbf{h}}_{MP}\|$ will reduce the power used for forwarding the residual self-interference and noises and thus improve the efficiency of power usage. However, when the ICI power $|\bar{h}_{Mk}|^2$ is very large, the impact of $\|\bar{\mathbf{h}}_{MP}\|$ can be neglected as shown in~(\ref{E:maxSINR-largeI}).

\item \emph{\textbf{Impact of ${\mathbf{h}}_{Pk}$ and $\bar{h}_{Mk}$}:} As shown by (\ref{E:SINR3-HD}), (\ref{E:maxSINR-largeS}) and (\ref{E:maxSINR-largeI}), increasing the desired signal $P_0\|\mathbf{h}_{Pk}\|^2$ can improve the performance of both the FD scheme and the HD scheme.\footnote{Herein, we consider that the increase of the desired signal's strength comes from reducing the distance between the PUE and its serving PBS, but not from increasing the power of the PBS. As a result, the interference from surrounding PBSs to the PUE will be even weaker, making it reasonable to focus on the dominant cross-tier ICI as we considered.} However, the performance improvement for the FD scheme is more significant because (\ref{E:FD-HD-1}) shows that the performance gain of FD over HD increases with $P_0\|\mathbf{h}_{Pk}\|^2$. It can also be seen that the performance gain of FD over HD decreases with the ICI power $|\bar{h}_{Mk}|^2$, until vanishes for very large $|\bar{h}_{Mk}|^2$ as shown by (\ref{E:maxSINR-largeI}). It should be pointed out that the extreme case with very strong ICI in (\ref{E:maxSINR-largeI}) rarely happens in practice even when the maximum CRE offset of $9$~dB in LTE systems is considered~\cite{CRE2013}. As will be verified in Fig.~\ref{F:placement}, the fICIC still exhibits evident performance gain over the HD scheme when the average power of the ICI is $10.4$~dB stronger than the desired signal.

%\item \emph{\textbf{Impact of FD transmission}:} When the ICI is large such that the PBS has no enough transmit power to thoroughly eliminate it through forwarding the listened ICI, the PBS needs to balance the power allocated to the desired signal and the forwarded ICI. Therefore, only part of power can be used for each of the two purposes. However, we can observe an appealing result from (\ref{E:maxSINR-largeI}) that the transmit power of the PBS is reusable in the sense that the PBS can use full power to transmit desired signal and at the same time use full power to reduce ICI (i.e., the term $P_0\|\mathbf{h}_{Pk}\|^2$ exists in both the nominator and the denominator of (\ref{E:maxSINR-largeI})). The seemingly contradictory results can be explained as follows. With the optimal $\beta^\star$ given in (\ref{E:w1-app}), we can compute the strength of the desired signal and ICI from (\ref{E:Objective1}) as $\kappa\cdot P_0\|\mathbf{h}_{Pk}\|^2$ and $\kappa^2\cdot |\bar{h}_{Mk}|^2$ with $\kappa = 1 - \frac{P_0\|\mathbf{h}_{Pk}\|^2}{|\bar{h}_{Mk}|^2}$, respectively, which results in the SINR shown in (\ref{E:maxSINR-largeI}). Since  $\kappa < 1$, we can find that the fICIC scheme sacrifices the strength of the desired signal to achieve more significant reduction of the ICI, which leads to the illusion that the power of the PBS is reusable.

\end{itemize}

\subsubsection{Large Residual Self-interference} When the self-interference cancellation for FD is imperfect and the residual self-interference, $\sigma_I^2$, is large, the parameter $B$ is large and the term $\frac{AC^2}{B}$ is small. Then by using the first-order taylor expansion, $\sqrt{c-z} \doteq \sqrt{c} - \frac{1}{2\sqrt{c}}z$ for small $z$, we can obtain $\beta^\star \doteq \frac{A}{(A+D)B}$ when $\frac{AC^2}{B}$ approaches to zero, % as
%\begin{equation}\label{E:w-large-selfI}
%  \beta^\star\doteq \frac{A}{(A+D)B}.
%\end{equation}
with which we can obtain from (\ref{E:maxSINR}) the maximal SINR of $\ms_k$~as
\begin{equation}\label{E:SINR-large-selfI}
  \widetilde{\mathtt{SINR}}_{k,FD}^\star \doteq \frac{P_0\|\mathbf{h}_{Pk}\|^2}{|\bar{h}_{Mk}|^2 + \sigma_n^2} = \mathtt{SINR}_{k,HD}^\star.
\end{equation}
Moreover, we can compute the power of the PBS allocated for forwarding the listened ICI from (\ref{E:Powerforward})  as
\begin{align} \label{E:Forward-power}
  %&\|\mathbf{W}_f\bar{\mathbf{h}}_{MP}\|^2 + (\sigma_I^2 + \sigma_n^2)\mathtt{tr}\left(\mathbf{W}_f^H\mathbf{W}_f\right) \nonumber\\
  & P_{out}^{fw}\doteq \frac{P_0^2\|\bar{\mathbf{h}}_{MP}\|^2|\bar{h}_{Mk}|^2\|\mathbf{h}_{Pk}\|^2}{(|\bar{h}_{Mk }|^2 \!+\! P_0\|\mathbf{h}_{Pk}\|^2 \!+ \! \sigma_n^2)^2(\|\bar{\mathbf{h}}_{MP}\|^2\! +\! \sigma_I^2 \!+\! \sigma_n^2)}.
\end{align}

It can be seen from (\ref{E:Forward-power}) that the transmit power of the PBS allocated for forwarding ICI decreases with the growth of residual self-interference $\sigma_I^2$. When $\sigma_I^2$ is very large, the PBS will use all power to transmit desired signals, and therefore the fICIC will reduce to the HD scheme.

\section{Narrowband Multi-user Case}
In this section, we consider the general multi-user case, where the MBS serves ${K_M}$ MUEs and the PBS serves ${K_P}$ PUEs with ${K_M} \geq 1$ and ${K_P} \geq 1$.

% \subsection{Optimal fICIC Scheme}
We optimize the fICIC scheme, aimed at maximizing the sum rate of ${K_P}$ PUEs served by the reference PBS while guaranteeing the fairness among the PUEs, subject to the maximal transmit power constraint of the PBS. The problem can be formulated as
\begin{subequations} \label{E:problem-MU}
    \begin{align}
        & \underset{\mathbf{W}_f, \{\mathbf{w}_{d,k}\}}{\max}\ R_{sum}  \label{E:Objective-MU}\\
        & s.\ t.\ \log(1 \!+\! \mathtt{SINR}_{k,FD}) \!=\! \alpha_k R_{sum}, \ k = 1,\dots, K_P \label{E:fairness-constraint}\\
        & \ \ \ \  \mathtt{tr}\left(\bar{\mathbf{H}}_{MP}\mathbf{W}_f^H\mathbf{W}_f\bar{\mathbf{H}}_{MP}^H\right) + (\sigma_I^2 + \sigma_n^2)\mathtt{tr}\left(\mathbf{W}_f^H\mathbf{W}_f\right)\nonumber\\
        & \qquad\qquad\qquad\qquad\qquad\qquad + \sum_{k=1}^{K_P} \|\mathbf{w}_{d,k}\|^2 \leq P_0, \label{E:Constraint-MU}
    \end{align}
\end{subequations}
where $R_{sum}$ is the sum rate of all PUEs, the constraints in (\ref{E:fairness-constraint}) ensure the data rate proportion among the PUEs with predefined fairness factors $\{\alpha_k\}$ as considered in~\cite{Emil2012}, $\alpha_k \geq 0$, and $\sum_{k=1}^{K_P} \alpha_k = 1$.

The expression of $\mathtt{SINR}_{k,FD}$ in multi-user case is given in (\ref{E:SINR3}), which is a function of the transmit power $P_{out}$ as shown in (\ref{E:Pout}) and thus is very complicated. To simplify $\mathtt{SINR}_{k,FD}$, we can prove that the optimal solution to problem (\ref{E:problem-MU}) is obtained when the power constraint (\ref{E:Constraint-MU}) holds with equality.\footnote{Otherwise, suppose that the power constraint holds with inequality with the optimal precoders $\mathbf{W}^\star_{f}$ and $\{\mathbf{w}_{d,k}^\star\}$, then given $\mathbf{W}^\star_{f}$, we can always find new precoders $\{\mathbf{w}'^\star_{d,k}\}$, defined as $\mathbf{w}'^\star_{d,k} =
c\mathbf{w}^\star_{d,k}$ for $k = 1, \dots, {K_P}$ with $c > 1$, which can further improve the SINR of all PUEs until the constraint (\ref{E:Constraint-MU}) holds with equality.} Based on this result, we have $P_{out} = P_0$ in (\ref{E:SINR3}).

Problem (\ref{E:problem-MU}) is to find the maximal achievable $R_{sum}$, denoted by $R_{sum}^\star$, ensuring all constraints satisfied, which can be solved by bisection methods~\cite{boyd2009convex}. Specifically, for a given $R_{sum}$ in an iteration, denoted by~$R_{sum}^0$, if the following optimization problem
\begin{subequations} \label{E:problem-MU-bisection}
    \begin{align}
        \mathtt{Find}\ & \mathbf{W}_f, \{\mathbf{w}_{d,k}\}  \label{E:Objective-MU-bisection}\\
         s.\ t.\ \  &  (\ref{E:Constraint-MU})\nonumber \\
         & \mathtt{SINR}_{k,FD} = 2^{\alpha_k R_{sum}^0} - 1, \ \ k = 1,\dots, K_P \label{E:fairness-constraint-bisection}
    \end{align}
\end{subequations}
is feasible, then it follows that $R_{sum}^0$is an achievable sum rate of all ${K_P}$ $\ms$s, i.e., $R_{sum}^0 \leq R_{sum}^\star$, otherwise, $R_{sum}^0 > R_{sum}^\star$. This condition can be used in bisection algorithms to find $R_{sum}^\star$.

Now the remaining issue is to find efficient approaches to evaluate the feasibility of problem (\ref{E:problem-MU-bisection}). In the following, we show that the feasibility problem can be solved by investigating the optimization problem below
\begin{subequations} \label{E:problem-MU-reform-feasibility}
    \begin{align}
        \underset{\mathbf{W}_f, \{\mathbf{w}_{d,k}\}}{\min}\ & \mathtt{tr}\left(\bar{\mathbf{H}}_{MP}\mathbf{W}_f^H\mathbf{W}_f\bar{\mathbf{H}}_{MP}^H\right) + (\sigma_I^2 + \sigma_n^2)\cdot\nonumber\\
        &\qquad\mathtt{tr}\left(\mathbf{W}_f^H\mathbf{W}_f\right) + \sum_{k=1}^{K_P} \|\mathbf{w}_{d,k}\|^2  \label{E:Objective-MU-reform-feasibility}\\
         s.\ t.\ \ &  \mathtt{SINR}_{k,FD} \geq \gamma_{k}, \ k = 1,\dots, {K_P}, \label{E:constraint-SINR-feasibility}
    \end{align}
\end{subequations}
where the objective function is the left-hand side of constraint (\ref{E:Constraint-MU}), and $\gamma_k \triangleq 2^{\alpha_k R_{sum}^0} - 1$.

To see this, we note that the optimal solution to problem (\ref{E:problem-MU-reform-feasibility}) is obtained when all SINR constraints in (\ref{E:constraint-SINR-feasibility}) hold with equality, otherwise, the value of the objective function can be always further reduced by properly decreasing $\|\mathbf{w}_{d,k}\|^2$. It suggests that if the minimum of the objective function (\ref{E:Objective-MU-reform-feasibility}) is smaller than $P_0$, then constraints (\ref{E:Constraint-MU}) and (\ref{E:fairness-constraint-bisection}) in problem (\ref{E:problem-MU-bisection}) are all satisfied. This means that  problem (\ref{E:problem-MU-bisection}) is feasible and $R_{sum}^0$ is an achievable sum rate, otherwise, problem (\ref{E:problem-MU-bisection}) will be infeasible. In what follows, we solve problem (\ref{E:problem-MU-reform-feasibility}).

By defining $\mathbf{w}_f = \mathtt{vec}(\mathbf{W}_f^H)$, we can rewrite problem (\ref{E:problem-MU-reform-feasibility})~as
\begin{subequations} \label{E:problem-MU-reform-feasibility-reform}
    \begin{align}
        & \underset{\mathbf{w}_f, \{\mathbf{w}_{d,k}\}}{\min} \|\big(\mathbf{I}_{N_t}\!\otimes\! \bar{\mathbf{H}}_{MP}\big)\! \mathbf{w}_f\|^2\!
        +\! (\!\sigma_I^2 + \sigma_n^2\!)\|\mathbf{w}_f\|^2 \!+\! \sum_{k=1}^{K_P}\! \|\mathbf{w}_{d,k}\|^2  \label{E:Objective-MU-reform-feasibility-reform}\\
         &s.\ t.\ \frac{|\mathbf{h}_{Pk}^H\mathbf{w}_{d,k}|^2}{\Omega_k} \geq \gamma_k, \ \forall k,
         .\label{E:constraint-SINR-feasibility-reform}
    \end{align}
\end{subequations}
where $\Omega_k \triangleq {\sum_{j\neq k}} \|\mathbf{h}_{Pk}^H \mathbf{w}_{d,j}\|^2 + \|\bar{\mathbf{h}}_{Mk} + e^{-j\phi} \left(\mathbf{h}_{Pk}^T \otimes\bar{\mathbf{H}}_{MP}\right) \mathbf{w}_f\|^2  + \|\left(\mathbf{h}_{Pk}^T\otimes\mathbf{I}_{N_r}\right) \mathbf{w}_f\|^2(\sigma_I^2+\sigma_n^2) + \sigma_n^2$.
%where the formula $\mathtt{vec}(\mathbf{AXB}) = (\mathbf{B}^T\otimes\mathbf{A})\mathtt{vec}(X)$ and $\mathtt{tr}(\mathbf{A^HB}) = \mathtt{vec}(A)^H\mathtt{vec}(B)$ are used, and $\otimes$ is the kronecker product.

Problem (\ref{E:problem-MU-reform-feasibility-reform}) is non-convex because the constraints in (\ref{E:constraint-SINR-feasibility-reform}) are non-convex. A common method to solve such a problem is to convert the non-convex SINR constraints into convex second-order cone constraints~\cite{Yu07}. Specifically, since
adding any phase rotation to $\mathbf{w}_{d,k}$ will not affect the SINR of all $\ms$s, we can assume that $\mathbf{h}_{Pk}^H \mathbf{w}_{d,k}$ is real-valued, which does not affect the global optima of problem (\ref{E:problem-MU-reform-feasibility-reform}). Then we can convert the non-convex problem (\ref{E:problem-MU-reform-feasibility-reform}) into the following second-order cone constrained problem
\begin{subequations} \label{E:problem-MU-reform-feasibility-reform-socp}
    \begin{align}
        \underset{\mathbf{w}_f, \mathbf{w}_{d}}{\min} & \|\big(\mathbf{I}_{N_t}\otimes \bar{\mathbf{H}}_{MP}\big)\! \mathbf{w}_f\|^2
        \!+\! (\sigma_I^2 \!+\! \sigma_n^2)\!\|\mathbf{w}_f\|^2 \!+\! \|\mathbf{w}_{d}\|^2  \label{E:Objective-MU-reform-feasibility-reform-socp}\\
         s.\ t.\ &  \sqrt{1 + \frac{1}{\gamma_k}} \mathbf{h}_{Pk}^H\mathbf{w}_{d,k} \nonumber\\
         & \geq \left\| \left[\!
         \begin{smallmatrix}
            \left(\mathbf{I}_{N_r}\otimes \mathbf{h}_{Pk}^H\right) \mathbf{w}_d\\
            e^{-j\phi} \left(\mathbf{h}_{Pk}^T \otimes\bar{\mathbf{H}}_{MP}\right) \mathbf{w}_f\\
            \sqrt{\sigma_I^2+\sigma_n^2}\left(\mathbf{h}_{Pk}^T\otimes\mathbf{I}_{N_r}\right)\mathbf{w}_f\\
            \sigma_n
         \end{smallmatrix} \!\right]
         \!+\! \left[\!
            \begin{smallmatrix}
                \bar{\mathbf{0}}_{N_r}\\
                \bar{\mathbf{h}}_{Mk}\\
                \bar{\mathbf{0}}_{N_r}\\
                0
             \end{smallmatrix}
         \!\right] \right\|, \forall k, \label{E:constraint-SINR-feasibility-reform-socp}
    \end{align}
\end{subequations}
where $\mathbf{w}_d = \mathtt{vec}\left([\mathbf{w}_{d,1}, \dots, \mathbf{w}_{d,{K_P}}]\right)$.

The resultant problem (\ref{E:problem-MU-reform-feasibility-reform-socp}) is convex since both the objective function and constraints are convex, which can be solved by standard convex optimization algorithms~\cite{boyd2009convex}. However, the computational complexity of the standard algorithms is still too high and prohibits the practical use of the fICIC scheme, especially when the numbers of MUEs, PUEs, and the transmit and receive antennas at the PBS are large. In the following we strive to propose a low-complexity algorithm to solve problem (\ref{E:problem-MU-reform-feasibility-reform}), with which the optimal precoders are obtained with explicit expressions. We begin with the discussion regarding the strong duality of problem~(\ref{E:problem-MU-reform-feasibility-reform}).

Recalling that we have shown the equivalence between problem (\ref{E:problem-MU-reform-feasibility-reform}) and problem (\ref{E:problem-MU-reform-feasibility-reform-socp}), and also known that strong duality holds for problem (\ref{E:problem-MU-reform-feasibility-reform-socp}) because it is a convex problem. Further, along the lines of~\cite[App.A]{Yu07}, we can show the equivalence between the Lagrangian functions of problem (\ref{E:problem-MU-reform-feasibility-reform-socp}) and problem (\ref{E:problem-MU-reform-feasibility-reform}). Therefore, strong duality holds for the non-convex optimization problem (\ref{E:problem-MU-reform-feasibility-reform}).
This means that we can solve problem (\ref{E:problem-MU-reform-feasibility-reform}) with the Lagrange dual method.

The dual problem to problem (\ref{E:problem-MU-reform-feasibility-reform}) can be expressed as
\begin{subequations} \label{E:dual}
    \begin{align}
  \underset{\bar{\lambda}_k}{\max}\ & \underset{\mathbf{w}_f, \{\mathbf{w}_{d,k}\}}{\min} \ J(\bar{\lambda_k}, \mathbf{w}_f, \mathbf{w}_{d,k}) \\
  s.t. \ \ & \bar{\lambda}_k \geq 0, \ \forall k,
\end{align}
\end{subequations}
where $\bar{\lambda}_k$ is the lagrangian multiplier, and $J(\bar{\lambda_k}, \mathbf{w}_f, \mathbf{w}_{d,k})$ is the Lagrangian function of problem (\ref{E:problem-MU-reform-feasibility-reform}) with the expression
\begin{align} \label{E:Largrangian-reform}
  &{J(\bar{\lambda}_k, \mathbf{w}_f, \mathbf{w}_{d,k}) =}
  %=  \|\left(\mathbf{I}_{N_t}\otimes \bar{\mathbf{H}}_{MP}\right) \mathbf{w}_f\|^2
  %      + (\sigma_I^2 + \sigma_n^2)\|\mathbf{w}_f\|^2 + \sum_{k=1}^{K_P} \|\mathbf{w}_{d,k}\|^2 \nonumber\\
  %      & + \sum_{k=1}^{K_P} \lambda_k \left(\sum_{j\neq k} \|\mathbf{h}_{Pk}^H \mathbf{w}_{d,j}\|^2 + \|\bar{\mathbf{h}}_{Mk} + e^{j\phi} \left(\mathbf{h}_{Pk}^T \otimes\bar{\mathbf{H}}_{MP}\right) \mathbf{w}_f\|^2  + |\left(\mathbf{h}_{Pk}^T\otimes\mathbf{I}\right) \mathbf{w}_f|^2(P_0\sigma_e^2+\sigma_n^2) + \sigma_n^2 - \frac{|\mathbf{h}_{Pk}^H\mathbf{w}_{d,k}|^2}{\gamma_0}\right)\nonumber\\
{\left\|\left(\mathbf{I}_{N_t}\otimes \bar{\mathbf{H}}_{MP}\right) \mathbf{w}_f\right\|^2 + (\sigma_I^2 + \sigma_n^2)\|\mathbf{w}_f\|^2} \nonumber\\
     & { + \sum_{k=1}^{K_P} \bar{\lambda}_k \left(\|\bar{\mathbf{h}}_{Mk} + e^{-j\phi} \left(\mathbf{h}_{Pk}^T \otimes\bar{\mathbf{H}}_{MP}\right) \mathbf{w}_f\|^2 \right. }\nonumber\\
     & {\left.+ \|\left(\mathbf{h}_{Pk}^T\otimes\mathbf{I}_{N_r}\right) \mathbf{w}_f\|^2(\sigma_I^2+\sigma_n^2) + \sigma_n^2\right)} \nonumber\\
     & { \!+\! \sum_{k=1}^{K_P} \mathbf{w}^H_{d,k}\Big(\mathbf{I}_{N_t} \!+\!\sum_{j\neq k} \bar{\lambda}_j \mathbf{h}_{Pj}\mathbf{h}_{Pj}^H \!- \! \frac{\bar{\lambda}_k}{\gamma_k} \mathbf{h}_{Pk} \mathbf{h}_{Pk}^H\Big)\mathbf{w}_{d,k}.}
\end{align}

\subsection{Optimal Solution to $\bar{\lambda}_k$}
It can be seen from (\ref{E:Largrangian-reform}) that $\underset{\mathbf{w}_f, \{\mathbf{w}_{d,k}\}}{\min} \ J(\bar{\lambda_k}, \mathbf{w}_f, \mathbf{w}_{d,k}) \to -\infty$ except when $\mathbf{I}_{N_t} +\sum_{j\neq k} \bar{\lambda}_j \mathbf{h}_{Pj}\mathbf{h}_{Pj}^H - \frac{\bar{\lambda}_k}{\gamma_0} \mathbf{h}_{Pk} \mathbf{h}_{Pk}^H \succeq \mathbf{0}_{N_t}$. Hence, the dual problem (\ref{E:dual}) is equivalent to
\begin{subequations} \label{E:problem-dual}
    \begin{align}
        \underset{\mathbf{w}_f}{\min}\ & \underset{\bar{\lambda}_k}{\max}\ J(\bar{\lambda}_k, \mathbf{w}_f, \bar{\mathbf{0}}_{N_t}) \label{E:Objective-dual}\\
         s.\ t.\ &  \mathbf{I}_{N_t} +\sum_{j\neq k} \bar{\lambda}_j \mathbf{h}_{Pj}\mathbf{h}_{Pj}^H - \frac{\bar{\lambda}_k}{\gamma_k} \mathbf{h}_{Pk} \mathbf{h}_{Pk}^H \succeq \mathbf{0}_{N_t}, \ \forall k \label{E:con-lambda} \\
         & \bar{\lambda}_k \geq 0, \ \forall k,
    \end{align}
\end{subequations}
where the objective function (\ref{E:Objective-dual}) comes from the fact that $J(\bar{\lambda_k}, \mathbf{w}_f, \mathbf{w}_{d,k})$ given in (\ref{E:Largrangian-reform}) is minimized when $\mathbf{w}_{d,k} \!=\! \bar{\mathbf{0}}_{N_t}$.

It is difficult to directly find the optimal $\bar{\lambda}_k$ from problem (\ref{E:problem-dual}) due to the complicated semi-definite positive constraints (\ref{E:con-lambda}). To solve the problem, we simplify the constraints as follows.

\emph{\textbf{Proposition 2:}} The semi-definite positive constraints (\ref{E:con-lambda}) can be equivalently expressed as
\begin{equation} \label{E:Con-semidefinite}
  \bar{\lambda}_k\mathbf{h}_{Pk}^H\Big(\mathbf{I}_{N_t} +\sum_{j\neq k} \bar{\lambda}_j \mathbf{h}_{Pj}\mathbf{h}_{Pj}^H\Big)^{-1} \mathbf{h}_{Pk} \leq \gamma_k,  \forall k.
\end{equation}
\begin{proof}
  See Appendix \ref{S:prooflemma4}.
\end{proof}

Note that for any given $\mathbf{w}_f$, the objective function of problem (\ref{E:problem-dual}) is an increasing function of $\bar{\lambda}_k$. Then we can show that the optimal $\bar{\lambda}_k$ maximizing $J(\bar{\lambda}_k, \mathbf{w}_f, \bar{\mathbf{0}}_{N_t})$ is obtained when the constraints in (\ref{E:Con-semidefinite}) hold with equality (otherwise, one can always increase $\bar{\lambda}_k$ to improve the value of the objective function). It suggests that the optimal $\bar{\lambda}_k^\star$ should satisfy
\begin{equation} \label{E:opt-lambda}
  \bar{\lambda}_k^\star = \frac{\gamma_k}{\mathbf{h}_{Pk}^H\big(\mathbf{I}_{N_t} +\sum_{j\neq k} \bar{\lambda}_j^\star \mathbf{h}_{Pj}\mathbf{h}_{Pj}^H\big)^{-1} \mathbf{h}_{Pk}}.
\end{equation}

(\ref{E:opt-lambda}) provides a fixed-point iterative algorithm to find the optimal $\bar{\lambda}_k^\star$, which can be expressed~as
\begin{equation} \label{E:opt-lambda11}
  \bar{\lambda}_k^{\star(n+1)} = \frac{\gamma_k}{\mathbf{h}_{Pk}^H\big(\mathbf{I}_{N_t} +\sum_{j\neq k} \bar{\lambda}_j^{\star(n)} \mathbf{h}_{Pj}\mathbf{h}_{Pj}^H\big)^{-1} \mathbf{h}_{Pk}}, %\ j, k = 1, \dots, {K_P},
\end{equation}
where the superscript $(n)$ denotes the $n$-th iteration.

The convergence of the fixed-point iterative algorithm can be proved based on the standard function theory~\cite{Wiesel2006}, which shows that the algorithm given in (\ref{E:opt-lambda11}) will converge to a unique optimal solution from any initial value $\{\bar{\lambda}_k^{\star(0)}\}$.

%KKT condition for $\mathbf{w}_f$:
%\begin{align}
%  &J =  \|\left(\mathbf{I}\otimes \bar{\mathbf{H}}_{MP}\right) \mathbf{w}_f\|^2
%        + (P_0\sigma_e^2 + \sigma_n^2)\|\mathbf{w}_f\|^2 + \sum_{k=1}^{K_P} \|\mathbf{w}_{d,k}\|^2 \nonumber\\
%        & + \sum_{k=1}^{K_P} \lambda_k \left(\sum_{j\neq k} \|\mathbf{h}_{Pk}^H \mathbf{w}_{d,j}\|^2 + \|\bar{\mathbf{h}}_{Mk} + e^{j\phi} \left(\mathbf{h}_{Pk}^T \otimes\bar{\mathbf{H}}_{MP}\right) \mathbf{w}_f\|^2  + |\left(\mathbf{h}_{Pk}^T\otimes\mathbf{I}\right) \mathbf{w}_f|^2(P_0\sigma_e^2+\sigma_n^2) + \sigma_n^2 - \frac{|\mathbf{h}_{Pk}^H\mathbf{w}_{d,k}|^2}{\gamma_k}\right).
%\end{align}

\subsection{Optimal Solution to $\mathbf{w}_f$}
After obtaining the optimal $\bar{\lambda}_k^\star$ for $k = 1,\dots, {K_P}$, we next find the optimal $\mathbf{w}_f^\star$ based on the Karush-Kuhn-Tucker (KKT) condition~\cite{boyd2009convex} associated with the Lagrangian function $J(\bar{\lambda}_k^\star, \mathbf{w}_f, \mathbf{w}_{d,k})$ . We can obtain that the optimal $\mathbf{w}_{f}^\star$ satisfies
\begin{align} \label{E:optimalwf}
  &\sum_{k=1}^{K_P} e^{j\phi} \bar{\lambda}_k^\star \left(\mathbf{h}_{Pk}^T \otimes\bar{\mathbf{H}}_{MP}\right)^H \bar{\mathbf{h}}_{Mk} + \left(\mathbf{I}_{N_t}\otimes \bar{\mathbf{H}}_{MP}^H\bar{\mathbf{H}}_{MP}\right) \mathbf{w}_f^\star \nonumber\\
  & + (\sigma_I^2+\sigma_n^2)\mathbf{w}_f^\star
  + \sum_{k=1}^{K_P} \bar{\lambda}_k^\star\left( \left((\mathbf{h}_{Pk}\mathbf{h}_{Pk}^H)^T \otimes\bar{\mathbf{H}}_{MP}^H\bar{\mathbf{H}}_{MP}\right)\mathbf{w}_f^\star \right.   \nonumber\\
  &\left. + (\sigma_I^2+\sigma_n^2) \left((\mathbf{h}_{Pk}\mathbf{h}_{Pk}^H)^T\otimes\mathbf{I}_{N_r}\right)\mathbf{w}_f^\star\right) = \bar{\mathbf{0}}_{N_tN_r}.
\end{align}

By applying the properties of Kronecker product~\cite{meyer2000matrix} and after some manipulations,
we can obtain the optimal $\mathbf{w}_f^\star$ as
\begin{align} \label{E:optimalwf2}
  & \mathbf{w}_f^\star  \!=\! \!-\!e^{j\phi} \big( \big(\sum_{k=1}^{K_P} \bar{\lambda}_k^\star(\mathbf{h}_{Pk}\mathbf{h}_{Pk}^H)^T \!+\! \mathbf{I}_{N_t}\big)^{-\!1} \big(\sum_{k=1}^{K_P} \bar{\lambda}_k^\star\bar{\mathbf{h}}_{Mk} \mathbf{h}_{Pk}^H\big)^T  \nonumber\\
  & \otimes \!\big(\bar{\mathbf{H}}_{MP}^H\bar{\mathbf{H}}_{MP}\!+\! (\sigma_I^2 \!+\! \sigma_n^2)\mathbf{I}_{N_r}\big)^{-\!1}\bar{\mathbf{H}}_{MP}^H\big)  \mathtt{vec}(\mathbf{I}_{N_tN_r}).
\end{align}

%\begin{align}
%  \mathbf{w}_f = & -e^{-j\phi} \left( \left(\sum_{k=1}^{K_P} \lambda_k(\mathbf{h}_{Pk}\mathbf{h}_{Pk}^H)^T + \mathbf{I}\right) \otimes \left(\bar{\mathbf{H}}_{MP}^H\bar{\mathbf{H}}_{MP}+ \Delta_2\mathbf{I}\right)\right)^{-1} \left(\left(\sum_{k=1}^{K_P} \lambda_k\bar{\mathbf{h}}_{Mk} \mathbf{h}_{Pk}^H\right)^T\otimes \bar{\mathbf{H}}_{MP}^H\right)\mathtt{vec}(\mathbf{I})\nonumber\\
%  = &  -e^{-j\phi} \left( \left(\sum_{k=1}^{K_P} \lambda_k(\mathbf{h}_{Pk}\mathbf{h}_{Pk}^H)^T + \mathbf{I}\right)^{-1} \left(\sum_{k=1}^{K_P} \lambda_k\bar{\mathbf{h}}_{Mk} \mathbf{h}_{Pk}^H\right)^T \otimes \left(\bar{\mathbf{H}}_{MP}^H\bar{\mathbf{H}}_{MP}+ \Delta_2\mathbf{I}\right)^{-1}\bar{\mathbf{H}}_{MP}^H\right)  \mathtt{vec}(\mathbf{I})
%\end{align}

Then we can recover the optimal $\mathbf{W}_f^\star$ from $\mathbf{w}_f^\star$ as
%\begin{align}
%  \mathbf{W}_f^H = & -e^{-j\phi} \left(\bar{\mathbf{H}}_{MP}^H\bar{\mathbf{H}}_{MP}+ \Delta_2\mathbf{I}\right)^{-1}\bar{\mathbf{H}}_{MP}^H \cdot \left(\sum_{k=1}^{K_P} \lambda_k\bar{\mathbf{h}}_{Mk} \mathbf{h}_{Pk}^H\right)\left(\sum_{k=1}^{K_P} \lambda_k\mathbf{h}_{Pk}\mathbf{h}_{Pk}^H + \mathbf{I}\right)^{-1}
%\end{align}
\begin{align} \label{E:Wf}
  &{\mathbf{W}_f^\star = } -e^{j\phi}\Big(\sum_{k=1}^{K_P} \bar{\lambda}_k^\star\mathbf{h}_{Pk}\mathbf{h}_{Pk}^H + \mathbf{I}_{N_t}\Big)^{-1} \cdot\\
  & \Big(\sum_{k=1}^{K_P} \bar{\lambda}_k^\star{\mathbf{h}}_{Pk} \bar{\mathbf{h}}_{Mk}^H\Big) \bar{\mathbf{H}}_{MP}\big(\bar{\mathbf{H}}_{MP}^H\bar{\mathbf{H}}_{MP}+ (\sigma_I^2+\sigma_n^2)\mathbf{I}_{N_r}\big)^{-1}.\nonumber
\end{align}

\subsection{Optimal Solution to $\mathbf{w}_{d,k}$ }

With the optimal $\lambda_k^\star$ and $\mathbf{W}_f^\star$, the optimal $\mathbf{w}_{d,k}^\star$ of problem (\ref{E:problem-MU-reform-feasibility-reform}) can be solved as follows.

According to the KKT condition, the optimal $\mathbf{w}_{d,k}^\star$ satisfies
%\begin{align}
%  &J =  \|\left(\mathbf{I}\otimes \bar{\mathbf{H}}_{MP}\right) \mathbf{w}_f\|^2
%        + (P_0\sigma_e^2 + \sigma_n^2)\|\mathbf{w}_f\|^2 + \sum_{k=1}^{K_P} \|\mathbf{w}_{d,k}\|^2 \nonumber\\
%        & + \sum_{k=1}^{K_P} \lambda_k \left(\sum_{j\neq k} \|\mathbf{h}_{Pk}^H \mathbf{w}_{d,j}\|^2 + \|\bar{\mathbf{h}}_{Mk} + e^{j\phi} \left(\mathbf{h}_{Pk}^T \otimes\bar{\mathbf{H}}_{MP}\right) \mathbf{w}_f\|^2  + |\left(\mathbf{h}_{Pk}^T\otimes\mathbf{I}\right) \mathbf{w}_f|^2(P_0\sigma_e^2+\sigma_n^2) + \sigma_n^2 - \frac{|\mathbf{h}_{Pk}^H\mathbf{w}_{d,k}|^2}{\gamma_k}\right)
%\end{align}
\begin{align}
  & \frac{\partial J(\bar{\lambda_k}^\star, \mathbf{w}_f^\star, \mathbf{w}_{d,k})}{\partial \mathbf{w}_{d,k}}\\
  &= 2\Big(\mathbf{I}_{N_t} + \sum_{j\neq k} \bar{\lambda}_j^\star \mathbf{h}_{P,j}\mathbf{h}_{P,j}^H - \frac{\bar{\lambda}_k^\star}{\gamma_k} \mathbf{h}_{Pk}\mathbf{h}_{Pk}^H\Big)\mathbf{w}_{d,k} = \bar{\mathbf{0}}_{N_t},\nonumber
 % = & \left(\mathbf{I} + \sum_{j\neq k} \lambda_j \mathbf{h}_{P,j}\mathbf{h}_{P,j}^H\right)\mathbf{w}_{d,k} - \frac{\lambda_k\mathbf{h}_{Pk}^H\mathbf{w}_{d,k}}{\gamma_k} \mathbf{h}_{Pk} = \mathbf{0}.
\end{align}
from which we have
\begin{align} \label{E:opt-wtk}
  \mathbf{w}_{d,k}^\star = & \frac{\lambda_k^\star\mathbf{h}_{Pk}^H\mathbf{w}_{d,k}^\star}{\gamma_k} \big(\mathbf{I}_{N_t} + \sum_{j\neq k} \bar{\lambda}_j^\star \mathbf{h}_{P,j}\mathbf{h}_{P,j}^H\Big)^{-1} \mathbf{h}_{Pk} \nonumber\\
   \triangleq & \sqrt{p_k^\star} \tilde{\mathbf{w}}_{d,k}^\star,
\end{align}
where $\tilde{\mathbf{w}}_{d,k}^\star =  \big(\mathbf{I}_{N_t} + \sum_{j\neq k} \bar{\lambda}_j^\star \mathbf{h}_{P,j}\mathbf{h}_{P,j}^H\big)^{-1} \mathbf{h}_{Pk}$, and $p_k^\star$ is a scalar controlling the power allocated to $\ms_k$ for transmitting desired signals.

To find the optimal $\{p_k^\star\}$ in (\ref{E:opt-wtk}), recalling that the optimal solution to problem (\ref{E:problem-MU-reform-feasibility-reform}) is obtained when all SINR constraints in (\ref{E:constraint-SINR-feasibility-reform}) hold with equality if problem (\ref{E:problem-MU-reform-feasibility-reform}) is feasible for given \{$\gamma_k$\}, then we can obtain the following equations with respect to $p_k^\star$
\begin{align} \label{E:opt-p}
    & \frac{p_k^\star|\mathbf{h}_{Pk}^H\tilde{\mathbf{w}}_{d,k}^\star|^2}{ \sum_{j\neq k} p_j^\star\|\mathbf{h}_{Pk}^H \tilde{\mathbf{w}}_{t,j}^\star\|^2 \!+\! \zeta_k(\mathbf{W}_f^\star)} \!= \!\gamma_k,\  k \!=\! 1,\dots, {K_P},
\end{align}
where $\zeta_k(\mathbf{W}_f^\star) = \|\bar{\mathbf{h}}_{Mk}^H + \mathbf{h}_{Pk}^H \mathbf{W}_f^\star \bar{\mathbf{H}}_{MP}^H e^{-j\phi}\|^2  + |\mathbf{h}_{Pk}^H \mathbf{W}_f^\star|^2(\sigma_I^2+\sigma_n^2) + \sigma_n^2$.

%Let $\zeta_k(\mathbf{W}_f)$ denote the term $\|\bar{\mathbf{h}}_{Mk} + e^{-j\phi} \left(\mathbf{h}_{Pk}^T \otimes\bar{\mathbf{H}}_{MP}\right) \mathbf{w}_f\|^2  + |\left(\mathbf{h}_{Pk}^T\otimes\mathbf{I}\right) \mathbf{w}_f|^2(\sigma_I^2+\sigma_n^2) + \sigma_n^2$ in the $\mathtt{SINR}$ constraints in (\ref{E:constraint-SINR-feasibility-reform}). Then based on Lemma 4 and (\ref{E:opt-wtk}), we can formulate the following equations to solve $p_k$ as
%\begin{equation} \label{E:opt-p}
%     \frac{p_k|\mathbf{h}_{Pk}^H\tilde{\mathbf{w}}_{d,k}|^2}{ \sum_{j\neq k} p_j\|\mathbf{h}_{Pk}^H \tilde{\mathbf{w}}_{t,j}\|^2 + \zeta_k(\mathbf{W}_f)} = \gamma_k,\  k = 1,\dots, {K_P},
%\end{equation}

From the equations in (\ref{E:opt-p}), we can solve the optimal $\mathbf{p}^\star\triangleq [p_1^\star, \dots, p_{K_P}^\star]^T$ as
\begin{equation} \label{E:PA}
  \mathbf{p}^\star = \mathbf{M}^{-1} \boldsymbol{\zeta},
\end{equation}
where $\boldsymbol{\zeta} = [\zeta_1(\mathbf{W}_f^\star), \dots, \zeta_{K_P}(\mathbf{W}_f^\star)]^T$, and $\mathbf{M}\in\mathbb{C}^{{K_P}\times {K_P}}$ is defined as
\begin{equation}
  [\mathbf{M}]_{kj} = \left\{
      \begin{array}{ll}
        \frac{1}{\gamma_k}|\mathbf{h}_{Pk}^H\tilde{\mathbf{w}}_{d,k}^\star|^2 , & k = j,\\
         - |\mathbf{h}_{Pk}^H\tilde{\mathbf{w}}_{t,j}^\star|^2, & k \neq j.
      \end{array}
  \right.
\end{equation}

Finally, we summarize the proposed low-complexity algorithm to solve the original optimization problem (\ref{E:problem-MU}) in general multiuser case in Table~\ref{tab:Distributed-P-B-allocation}.

\begin{table}[h]
\caption{Low-complexity algorithm for multi-user fICIC}
\label{tab:Distributed-P-B-allocation}
\hrule
\begin{algorithm}{aaa}
{}
\textbf{Initialization:} Set $i=0$, $\underline{R_{sum}}=0$, and $\overline{R_{sum}} = \sum_{k=1}^{K_P} \log(1 +\mathtt{SINR}_k^{ub})$, where $\mathtt{SINR}_k^{ub} = \frac{P_0\|\mathbf{h}_{Pk}\|^2}{\sigma_n^2}$ is an upper bound of the SINR of $\ms_k$, which is achieved when the ICI disappears and only $\ms_k$ is served by the PBS.\\
\textbf{Bisection Iteration:} At the $i$-th iteration, set $i \leftarrow i + 1$.
        \begin{itemize}
            \item[1)] Compute $R_{sum}^0 = \frac{\underline{R_{sum}} + \overline{R_{sum}}}{2}$.
            \item[2)] Given $R_{sum}^0$, compute $\{\bar{\lambda}_k^\star\}$ with the fixed-point iterative algorithm given by (\ref{E:opt-lambda11}).
            \item[3)] Given $\{\bar{\lambda}_k^\star\}$, compute $\mathbf{W}_f^\star$ based on (\ref{E:Wf}).
            \item[4)] Given $R_{sum}^0$, $\{\bar{\lambda}_k^\star\}$ and $\mathbf{W}_f^\star$, compute $\{\tilde{\mathbf{w}}_{d,k}^\star\}$ and $\{p_k^\star\}$ based on (\ref{E:opt-wtk}) and (\ref{E:PA}), respectively, then obtain $\mathbf{w}_{d,k}^\star = \sqrt{p_k^\star}\tilde{\mathbf{w}}_{d,k}^\star$.
            \item[5)] Given $\mathbf{W}_f^\star$ and $\{\mathbf{w}_{d,k}^\star\}$, compute the value of the objective function of problem (\ref{E:problem-MU-reform-feasibility}), denoted by $P_{0}^{i}$.
            \item[6)] \textbf{If} $P_{0}^{i} \leq P_0$, let $\underline{R_{sum}} \leftarrow R_{sum}^0$, \textbf{otherwise}, let $\overline{R_{sum}} \leftarrow R_{sum}^0$.
        \end{itemize}\\
\textbf{Repeat:} Iterate step 2 until the required accuracy is reached, i.e., $ \overline{R_{sum}} - \underline{R_{sum}} \leq \varepsilon$, where $\varepsilon$ is a specific threshold.
\end{algorithm}
\hrule
\end{table}

\section{Generalization to wideband systems}
In previous sections the fICIC is optimized in narrowband systems, where all the channels are flat fading and therefore only single-tap forwarding precoder is designed. When considering wideband systems, frequency-selective fading channels should be taken into account and hence a multi-tap finite impulse response (FIR) forwarding precoder needs to be designed, which makes the design of the fICIC more involved.

In this section, we generalize the fICIC to OFDM systems. For the sake of notational simplicity, we consider single-user case, i.e., $K_M=1$ and $K_P=1$, which can be straightforwardly extended to multiuser orthogonal frequency division multi-accessing (OFDMA) systems due to the orthogonality between subcarriers. Following the previous narrowband-system definitions, let $\mathbf{h}_{Mk}(t)$ and $\mathbf{h}_{Pk}(t)$ denote the time-domain channels from the MBS and the PBS to $\ms_k$, $\mathbf{H}_{MP}(t)$ denote the time-domain channel from the MBS to the PBS, and $\mathbf{H}_{PP}(t)$~denote the time-domain self-interference channel of the FD~PBS. The corresponding frequency-domain channels on the $n$-th subcarrier of the four channels are denoted by $\mathbf{g}_{Mkn}$, $\mathbf{g}_{Pkn}$, $\mathbf{G}_{MPn}$, and $\mathbf{G}_{PPn}$, respectively.

Consider that the PBS employs a FIR precoder $\mathbf{W}_f(t) = \sum_{l=0}^{L-1} \mathbf{W}_{fl} \delta(t-lT_s)$ to forward the listened ICI, where $L$ is the order of the FIR precoder and $T_s$ is the sampling interval. The selection of $L$ needs to ensure that the  delay spread of the equivalent channel for the forwarded ICI, $\mathbf{h}_{Pk}(t)\odot\mathbf{W}_f(t)\odot\mathbf{H}_{MP}(t)\odot\delta(t-\tau)$, does not exceed the cyclic prefix (CP) of the OFDM system in order to maintain orthogonality between subcarriers. Let $\bar{\mathbf{W}}_{fn}$ denote the frequency response of $\mathbf{W}_f(t)$ on the $n$-th subcarrier.

Then, following the same derivations in Section~\ref{S:system_model} for narrowband systems, we can obtain the transmit power of the PBS on the $n$-th subcarrier as
\begin{align} \label{E:Pout-wbn}
  &P_{out,n} = \\
  &\frac{\mathtt{tr}\big(\bar{\mathbf{W}}_{fn} \bar{\mathbf{g}}_{MPn}\bar{\mathbf{g}}_{MPn}^H \bar{\mathbf{W}}_{fn}^H\big) \!+ \! \sigma_n^2\mathtt{tr}\big(\bar{\mathbf{W}}_{fn}\bar{\mathbf{W}}_{fn}^H\big)\! +\! \|\bar{\mathbf{w}}_{dn}\|^2}{1 - \sigma_e^2\mathtt{tr}\big(\bar{\mathbf{W}}_{fn}\bar{\mathbf{W}}_{fn}^H\big)},\nonumber
\end{align}
where $\bar{\mathbf{g}}_{MPn}=\mathbf{G}_{MPn}^H\bar{\mathbf{w}}_{Mn}$ is the equivalent frequency-domain channel from the MBS to the PBS, $\bar{\mathbf{w}}_{Mn}$ is the precoder at the MBS for the MUE on the $n$-th subcarrier, $\bar{\mathbf{w}}_{dn}$ is the precoder at the PBS for the desired signal of $\ms_k$, and $\sigma^2_e = \frac{\sigma_n^2}{P_{tr}} + 2\bar{\alpha}_{PP}(\mu_x+\mu_y)$ is the same as defined in (\ref{E:x22}).
With (\ref{E:Pout-wbn}), the total transmit power constraint for the PBS can be expressed as
\begin{equation} \label{E:PBPC-wb}
 P_{out} = \sum_{n=1}^N P_{out,n}  \leq P_0.
\end{equation}

Compared to the power constraint in (\ref{E:Pout}) for narrowband system, which can be converted into a convex constraint for the precoders as shown in (\ref{E:PBPC}), the constraint in (\ref{E:PBPC-wb}) for wideband system is much more complicated and is non-convex.

Similar to (\ref{E:SINR3}), we can obtain the SINR of $\ms_k$ on the $n$-th subcarrier as
\begin{align}\label{E:SINR3-wb}
  &\mathtt{SINR}_{kn,FD} \nonumber\\
  &= |\mathbf{g}_{Pkn}^H\bar{\mathbf{w}}_{dn}|^2/\big(\|\bar{g}_{Mkn}^\ast + \mathbf{g}_{Pkn}^H \bar{\mathbf{W}}_{fn} \bar{\mathbf{g}}_{MPn} e^{-jdn}\|^2  \nonumber\\
  &\qquad + \|\mathbf{g}_{Pkn}^H \bar{\mathbf{W}}_{fn}\|^2(P_{out,n}\sigma_e^2+\sigma_n^2) + \sigma_n^2\big),
\end{align}
where $\bar{g}_{Mkn} \triangleq \bar{\mathbf{w}}_{Mn}^H\mathbf{g}_{Mkn}$ is the equivalent channel from the MBS to the $\ms$ on the $n$-th subcarrier, and
$P_{out,n}$ is given in~(\ref{E:Pout-wbn}), which is a function of the precoders $\bar{\mathbf{W}}_{fn}$ and $\bar{\mathbf{w}}_{dn}$.

Then, the wideband fICIC precoder optimization problem, aimed at maximizing the sum rate of $\ms_k$ over all subcarrier, can be formulated as
\begin{subequations} \label{E:problem-wb}
  \begin{align}
   & \max_{\mathbf{W}_f(t), \{\bar{\mathbf{w}}_{dn}\}} \sum_{n=1}^N \log(1 +\mathtt{SINR}_{kn,FD}) \label{E:objective-wb}\\
     & s.t. \big[[\bar{\mathbf{W}}_{f1}]_{ij}, \!\dots\!, [\bar{\mathbf{W}}_{fN}]_{ij}\big]^T\! =\! \mathbf{F}\big[[\mathbf{W}_{f1}]_{ij}, \!\dots\!, [\mathbf{W}_{fL}]_{ij}\big]^T  \label{E:cons-wb0}\\
    & \qquad \sum_{n=1}^N P_{out,n} \leq P_0, P_{out,n} \geq 0,\ \forall n, \label{E:cons-wb2}
  \end{align}
\end{subequations}
where constraint (\ref{E:cons-wb0}) restricts that the $N$ frequency-domain precoders $\{\bar{\mathbf{W}}_{fn}\}$ are generated from the $L$-tap time-domain FIR precoder $\mathbf{W}_f(t)$, and $\mathbf{F}\in\mathbb{C}^{N\times L}$ is the matrix containing the first $L$ columns of the $N\times N$ fast fourier transformation matrix.

Problem (\ref{E:problem-wb}) is non-convex, whose global optimal solution is difficult to find. We can obtain a local optimal solution to the problem by using a gradient-based solution (specifically using the function $\mathtt{fmincon}$ of the optimization toolbox of MATLAB). Note that the direction of $\bar{\mathbf{w}}_{dn}$ only affects the power of desired signal in the numerator of $\mathtt{SINR}_{n,FD}$. Therefore, the direction of the optimal $\bar{\mathbf{w}}_{dn}$ can be obtained as $\frac{\bar{\mathbf{w}}_{dn}^\star}{\|\bar{\mathbf{w}}_{dn}^\star\|} = \frac{\mathbf{g}_{Pkn}}{\|\mathbf{g}_{Pkn}\|}$,
with which the number of variables in problem (\ref{E:problem-wb}) is reduced.

\section{Practical Issues}
By now, we have introduced the concept of fICIC and optimized the associated precoders. In this section, we discuss some practical issues regarding the application of the~fICIC.

\subsection{Channel Acquisition}
To apply the fICIC, the PBS needs to have the channels from the MBS to both the PBS and PUEs, i.e., ${\mathbf{H}}_{MP}$ and ${\mathbf{h}}_{Mk}$, as well as the channel from the PBS to PUEs, $\mathbf{h}_{Pk}$, $\forall k$. First, the channel ${\mathbf{H}}_{MP}$ can be directly estimated at the PBS by using the FD receive antennas to receive the downlink training signals sent from the MBS. Second, noting that in TDD systems channel reciprocity holds between the HD transceiver at the PBS and each PUE, the channel $\mathbf{h}_{Pk}$ can be estimated at the PBS by using the HD receive antennas to receive the uplink training signals sent by $\ms_k$. Finally, the channel ${\mathbf{h}}_{Mk}$ can be first estimated by $\ms_k$ and then fed back to the PBS, where digital or analog feedback schemes can be employed~\cite{Marzetta2006}. We will evaluate the performance of the fICIC under imperfect channel estimation and feedback in next section.

\subsection{fICIC {v.s.} the HD Scheme}
As we analyzed in Section \ref{S:asymptotic}, with the fICIC, the FD PBS can adaptively switch between FD mode and HD mode by optimizing the precoders for transmitting desired signals and forwarding listened interference. For instance, when the ICI is very strong or the residual self-interference is very large, the fICIC will reduce to the HD scheme, as shown by (\ref{E:maxSINR-largeI}) and (\ref{E:SINR-large-selfI}). Therefore, with perfect channel information at the PBS, the proposed fICIC will always outperform the HD scheme given the same number of uplink and downlink RF chains and BB modules. Yet, to support the FD technique extra passive antennas (though cheap) and FD modules are required. When imperfect channels are considered at the PBS, the performance of the fICIC will decrease. Nevertheless, simulations in next section show that significant performance gain can be still achieved by the fICIC over the HD scheme with imperfect channels.

\subsection{Self-interference Cancellation} \label{S:SIC}
Effective self-interference cancellation is crucial for the fICIC, where isolation of the transmit and FD receive antennas is an important approach~\cite{Everett2014}. Considering the transceiver structure of the FD PBS given in Fig.~\ref{F:system}, where the FD receive antennas are only active in the downlink to receive the ICI from the MBS while not receiving the uplink signals from the PUEs, the FD receive antennas can be mounted far away from the transmit antennas, e.g., installing the FD receive antennas outside a building and transmit antennas inside, respectively. In this manner, the self-interference can be largely suppressed in general.

\subsection{Joint Application with Existing ICIC Schemes}
Existing ICIC schemes including eICIC and CoMP-CB operate at the MBS, while the fICIC can operate at each PBS individually. Therefore, the fICIC can be directly applied together with existing ICIC schemes. As will be shown in next section, a joint application of the fICIC and existing schemes can achieve evident performance improvement.

\section{Simulation Results}
In this section we verify our analytical results and evaluate the performance of the proposed fICIC scheme via simulations.

\begin{figure}
\centering
\includegraphics[width=0.46\textwidth]{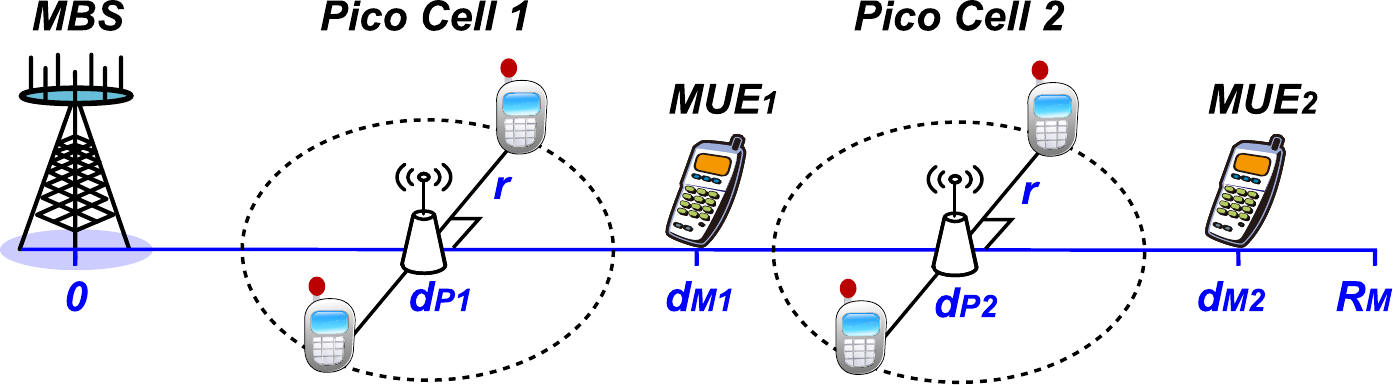}
\caption{Network layout for simulations, where two PBSs are deployed in the macro cell covered by the MBS.}\label{F:layout}
\end{figure}

\subsection{Simulation Setups}
For all simulations, unless otherwise specified, the parameters in this subsection are used.
The considered HetNet layout is shown in Fig.~\ref{F:layout}, where the MBS is located at the center of the macro cell, $\bs_1$ and $\bs_2$ are respectively located at $(d_{P1}, 0)$ and $(d_{P2}, 0)$, the MBS serves two MUEs located at $(d_{M1}, 0)$ and $(d_{M2}, 0)$, and $\bs_k$ serves two PUEs located at $(d_{Pk}, r)$ and $(d_{Pk}, -r)$, respectively. We set the radius of the macro cell $R_M$ as $500$~$m$, $d_{P1} = 60$~$m$, $d_{P2} = 180$~$m$, $r = 40$~$m$, $d_{M1} = 120$~$m$, and $d_{M2}= 240$~$m$. Since we have shown in the analytical results that the performance of the fICIC depends on the strength of the ICI, the positions of the two PBSs, $d_{P1}$ and $d_{P2}$, are selected to reflect the cases of strong and weak ICI, respectively. We will also evaluate the performance of the fICIC with random PBS placements later.

The MBS transmits with $M = 4$ antennas and the power of $P_M = 46$~dBm, each PBS has $N_r = 2$ FD receive antennas and transmits with $N_t = 2$ antennas and the maximal power of $P_0 = 30$~dBm, and each UE (including MUE and PUE) has one receive antenna. The path loss is set as  $128.1 + 37.6 \log_{10}d$ for macro cell and $140.7 + 36.7 \log_{10} d$ for pico cell, respectively, where $d$ is the distance in $km$~\cite{TR36.814}. A penetration loss of $20$~dB is considered for the channels to PUEs. We model the interference from the MBSs in adjacent macro cells and the surrounding PBSs as noises. Define the average receive SNR of a MUE located at the cell edge as $\text{SNR}_{\text{edge}}$, then the noise variance $\sigma_n^2$ can be obtained as $\sigma_n^2 = P_M - (128.1 + 37.6 \log_{10} R_M) - \text{SNR}_{\text{edge}}$ in dBm. To evaluate the impact of imperfect self-interference cancellation for FD, we define the signal to self-interference ratio as $\text{SIR}_{\text{self}} = P_0 - \sigma_I^2$ in dB in order to reflect the level of self-interference cancellation. The Rayleigh flat small-scale fading channels are considered. The fairness factors are set as $\alpha_k=\frac{1}{K_P}$ in multi-user case. All the results are averaged over 1000 channel realizations.

\subsection{Narrowband Single-user Case}

\begin{figure}
\centering
\includegraphics[width=0.46\textwidth]{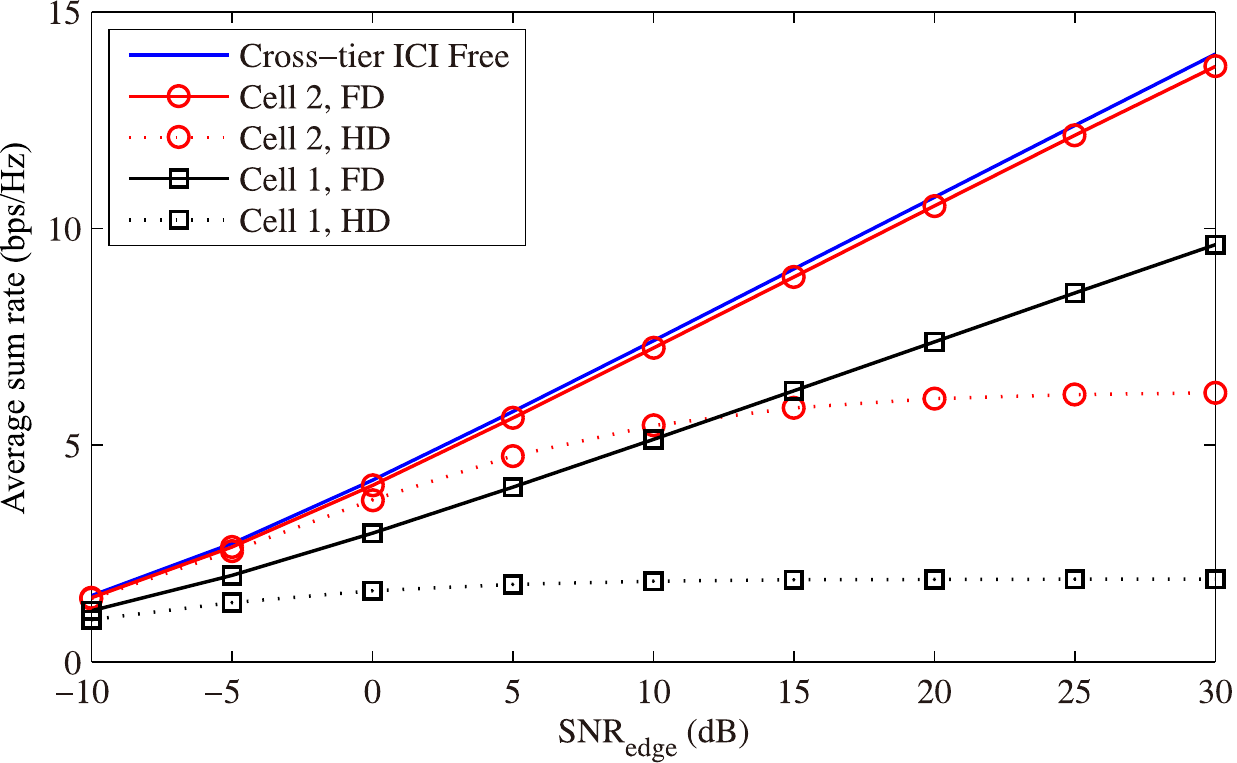}
\caption{Average sum rate versus $\text{SNR}_{\text{edge}}$, where perfect self-interference cancellation and perfect channels are considered, $K_M = 1$, $K_P = 1$, and $N_r = 1$. }\label{F:rate}
\end{figure}

%\begin{figure}
%\centering
%\includegraphics[width=0.7\textwidth]{Fig3_NtP2}
%\caption{Normalized residual ICI and the power of PBS allocated for ICI forwarding versus $\text{SNR}_{\text{edge}}$, where perfect self-interference cancellation and perfect channels are considered, $K_M = 1$, $K_P = 1$, and $N_r = 1$. }\label{F:PI}
%\vspace{-0.6cm}
%\end{figure}
We first simulate the single-user case, where the MBS serves one MUE, each PBS serves one PUE, and each PBS has $N_r = 1$ FD receive antenna.

Under the assumption of perfect self-interference cancellation, the average sum rate achieved by the fICIC is depicted in Fig.~\ref{F:rate}. For comparison, the performance of the HD scheme, given by $\mathcal{E}\{\log(1+\mathtt{SINR}_{k,HD}^\star)\}$, and the performance in ICI free case, given by $\mathcal{E}\{\log(1+\frac{P_0\|\mathbf{h}_{Pk}\|^2}{\sigma_n^2})\}$, are also presented. First, we can see that all the schemes perform closely in low SNR regime, where the system operates in noise-limited scenario. With the increase of SNR, the performance floor appears for the HD scheme, which is caused by the ICI from the MBS. The performance of Cell 2 is better than Cell 1 due to $d_{P2} > d_{P1}$ that leads to weaker ICI. The proposed fICIC exhibits a noticeable performance gain over the HD scheme. For weak ICI case, e.g., Cell 2, as we analyzed, the fICIC can thoroughly eliminate the ICI in a high probability (considering the randomness of small-scale channels), and thus the performance gap between the fICIC and the ICI free case is very small.

%To see more clearly how the fICIC suppresses the ICI, in Fig.~\ref{F:PI} we plot the normalized residual ICI achieved by the fICIC scheme, defined as $\frac{|\bar{h}_{Mk}^\ast + \mathbf{h}_{Pk}^H \mathbf{W}_f \bar{\mathbf{h}}_{MP} e^{-j\phi}|^2}{|\bar{h}_{Mk}|^2}$, as well as the power of each PBS allocated for forwarding the listened ICI, i.e., $P_{out}^{fw}$ given in (\ref{E:Powerforward}). We can find that the fICIC can thoroughly cancel the weak ICI experienced by Cell~2, while the strong ICI to Cell~1 cannot be fully canceled. Moreover, in order to compensate the stronger ICI, $\bs_1$ needs to allocate much larger power to forward the ICI, which reduces the power for transmitting desired signals, leading to performance degradation compared to Cell~2.

\begin{figure}
\centering
\includegraphics[width=0.46\textwidth]{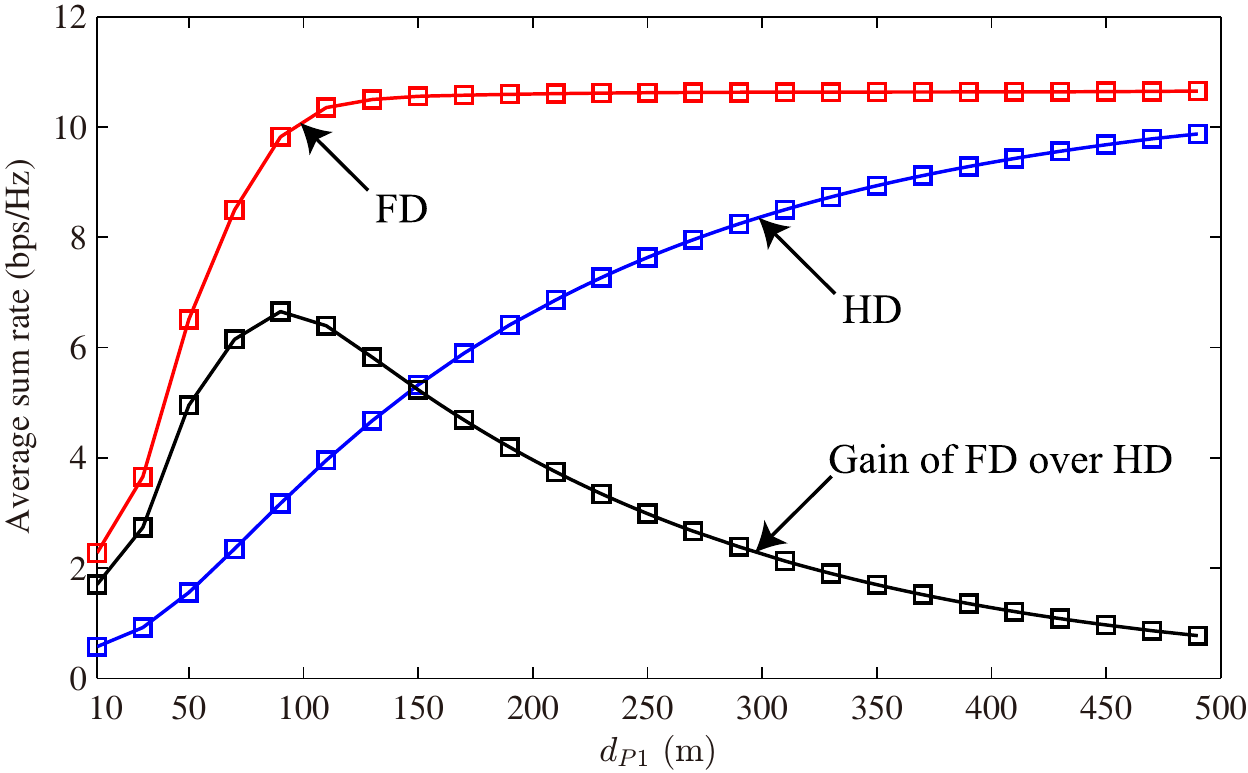}
\caption{Average sum rate versus $d_{P1}$ for different PBS placements, where perfect self-interference cancellation and perfect channels are considered, $K_M = 1$, $K_P = 1$, and $N_r = 1$. }\label{F:placement}
\end{figure}

In Fig.~\ref{F:placement}, we evaluate the impact of the strength of the ICI on the performance of the fICIC by simulating a single PBS with different positions $d_{P1}$. The performance achieved by the fICIC and the HD scheme as well as the performance gain of the fICIC are depicted. It is shown that the performance of the HD scheme increases slowly with the reduction of ICI, i.e., the increase of $d_{P1}$, while the performance of the fICIC first increases fast and then keeps nearly constant. It implies that the fICIC can make a large area of the macro cell, e.g., from $150$~$m$ to $500$~$m$, experience very weak ICI. Moreover, we can see that the fICIC performs close to the HD scheme at large $d_{P1}$ because the ICI is very weak and has negligible impact on the performance in this case. Further recalling that the fICIC will degenerate to the HD scheme when the ICI is very strong as analyzed in Section \ref{S:asymptotic}, we can understand that the gain of the fICIC over the HD scheme first increases and then decreases with~$d_{P1}$. However, we can still observe an evident performance improvement of nearly $300\%$ even when $d_{P1} = 10$~$m$, in which case the average power of the ICI is $10.4$~dB stronger than that of the desired signals.

\begin{figure}
\centering
\includegraphics[width=0.46\textwidth]{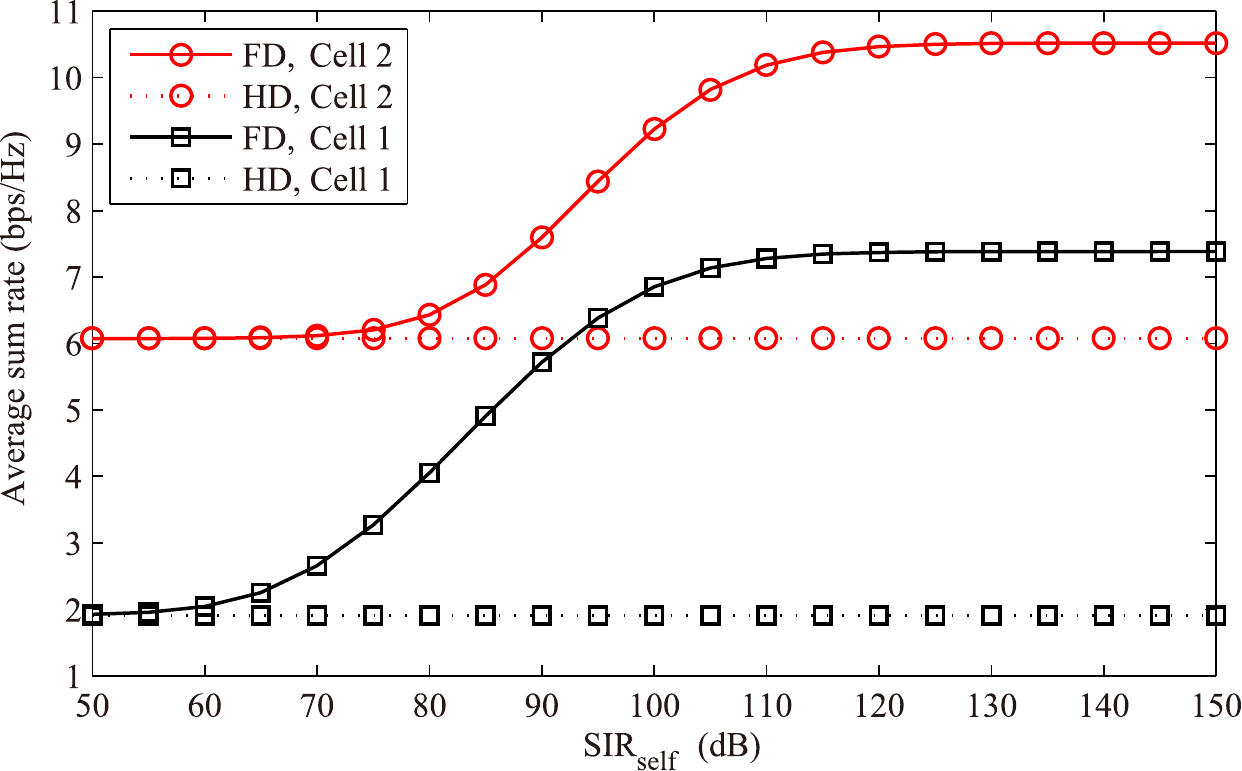}
\caption{Average sum rate versus $\text{SIR}_{\text{self}}$ with imperfect self-interference cancellation, where perfect channels are considered, $\text{SNR}_{\text{edge}} = 20$~dB, $K_M = 1$, $K_P = 1$, and $N_r = 1$.}\label{F:SIR}
\end{figure}

Figure~\ref{F:SIR} shows the performance of the fICIC as a function of $\text{SIR}_{\text{self}}$, where imperfect self-interference cancellation is considered. First, it can be seen that the fICIC reduces to the HD scheme for low $\text{SIR}_{\text{self}}$, which agrees with our previous analysis. Second, the fICIC outperforms the HD scheme when $\text{SIR}_{\text{self}} \geq 60$~dB for Cell 1 but $75$~dB for Cell 2. This is because $\bs_1$ is closer to the MBS and hence can listen a stronger ICI, which can relax the requirement of self-interference cancellation for FD. Moreover, it is shown that the self-interference cancellation of $110$~dB is sufficient for the fICIC, which is practically possible because on one hand existing work has reported $90$~dB self-interference cancellation even for closely placed transmit and receive antennas~\cite{Everett2014}, and on the other hand, as we discussed before, in our case the transmit antennas and FD receive antennas can be separated far apart, leading to further reduction of the self-interference, e.g., with an additional $20$~dB penetration loss.

\subsection{Narrowband Multi-user Case}

In this subsection the multi-user case is simulated, where each PBS serves two PUEs. We compare the performance of the fICIC with the HD scheme, a successive interference cancellation (SIC) based non-linear HD scheme (denoted by HD-SIC)~\cite{Wildemeersch2014}, the ABS-based eICIC and CoMP-CB, and also evaluate the performance of the combinations of the fICIC with eICIC and CoMP-CB. To study the effectiveness of the seven schemes for cancelling different-strength ICI, we fix the location of Cell 2 and adjust the location of Cell 1 to generate different interference to noise ratio (INR) for PUEs in Cell 1, where $\mathtt{INR}\triangleq\frac{\mathcal{E}\{\|\bar{\mathbf{h}}_{Mk}\|^2\}}{\sigma_n^2}$.
To enable the HD-SIC scheme, we consider that the MBS serves a single MUE because each PUE has only one antenna so that only one interference signal can be decoded and cancelled.

\begin{figure}
\centering
\includegraphics[width=0.48\textwidth]{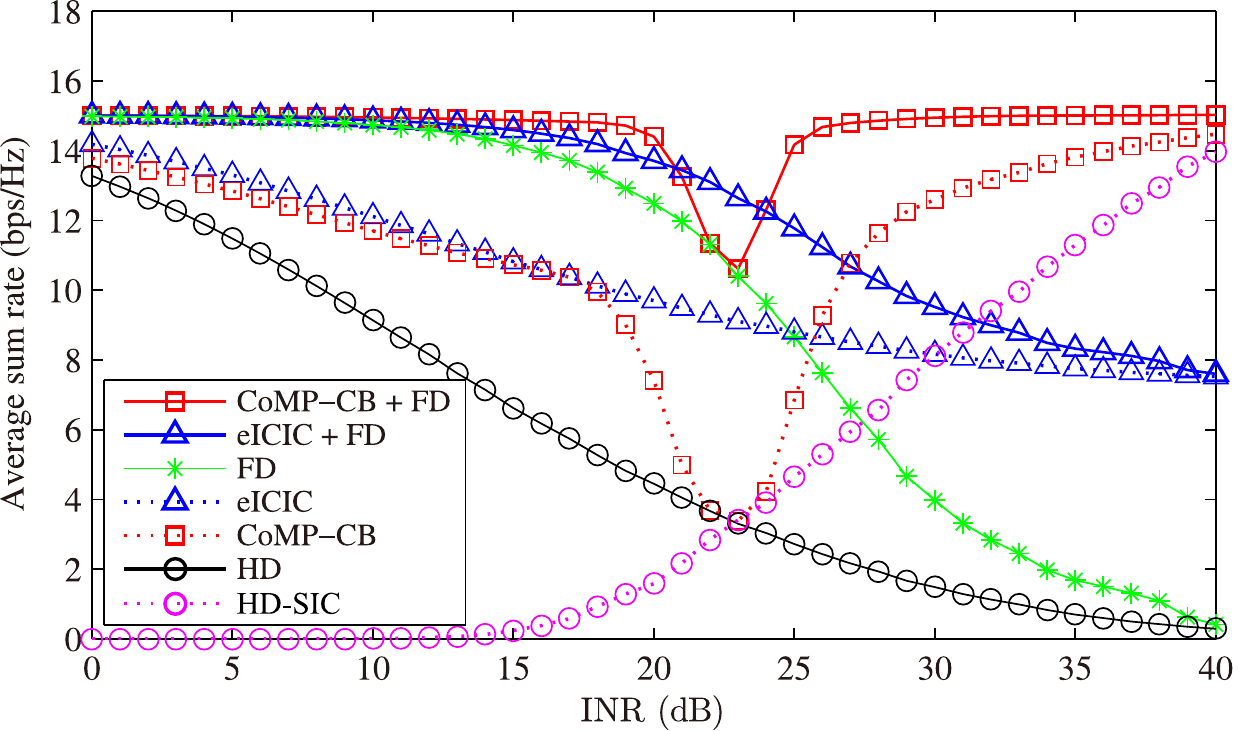}
\caption{Average sum rate of PUEs in Cell 1 achieved by relevant schemes versus $\text{INR}$, where perfect self-interference cancellation and perfect channels are considered, $\text{SNR}_{\text{edge}} = 20$~dB, $N_r = 2$, $K_M = 1$, and $K_P = 2$. }\label{F:eICIC}
\end{figure}

In the simulations, the proposed low-complexity algorithm given in Table~\ref{tab:Distributed-P-B-allocation} is employed to obtain the performance of both the fICIC and the HD scheme, where $\mathbf{W}_f$ is set as zero for the HD scheme. For the HD-SIC, since the ICI needs to be decoded first by regarding the desired signal from the PBS as interference, the PBS may not transmit with its maximal power. Given the data rate of the MUE as $4$~bps/Hz, we employ exhaustive searching to find the maximal feasible transmit power of the PBS under the ICI decoding constraint, where for any given transmit power the precoder of the PBS is computed the same as the HD scheme.
For the ABS-based eICIC, it is considered that the MBS mutes in half time to provide ICI-free environment to the PUEs. For CoMP-CB, the signal-to-leakage-plus-noise ratio (SLNR) based precoder~\cite{VSINR2009} is employed.
The results are depicted in Fig.~\ref{F:eICIC}.

Compared with the fICIC, we can see that the eICIC achieves worse performance when the ICI is not strong, say $\mathrm{INR}< 25$~dB, because the fICIC can effectively cancel the weak-medium level of ICI and allow the PUEs to use all the time-frequency resources, while with the eICIC the PUEs only uses half resources. When the ICI is strong, as we analyzed before, the fICIC is not effective, and the eICIC shows large performance gain. By combining eICIC with fICIC, the two schemes supplement each other and achieve much better performance.

For CoMP-CB, we can see that it is superior to the fICIC when the ICI is strong, say $\mathrm{INR} \geq 26$~dB, but inferior for weak-medium ICI. This can be explained as follows. In the simulated case, the MBS has only four antennas, which are not adequate to serve one MUE and suppress the ICI for four PUEs simultaneously. As a result, when the INR of Cell 1 is smaller than Cell 2, the SLNR-based CoMP-CB only suppresses the stronger ICI to the PUEs in Cell 2, while the ICI to PUEs in Cell 1 will be mitigated only for large INR (this also explains why the performance of Cell 1 with pure CoMP-CB or the combination of CoMP-CB and fICIC first deceases and then increases with the growth of INR.). By contrast, the fICIC is effective to suppress weak-medium ICI but not for strong ICI. Therefore, CoMP-CB and fICIC perform better in different ICI cases. Since CoMP-CB has no enough antenna resources to cancel weak-medium ICI in the simulation, the combination of CoMP-CB and fICIC has negligible gain over the pure fICIC for weak-medium INR. However, evident performance improvement is achieved by their combination for strong ICI, because the strong ICI can be first suppressed into weak-medium ICI by CoMP-CB and then further effectively cancelled by the fICIC.

For the HD-SIC, we can observe that it is not always feasible when the ICI is not strong, e.g., $\mathrm{INR} < 15$~dB. With the increase of ICI, the interference signal becomes decodable and more power can be transmitted by the PBS, which results in the performance improvement as expected. It can be seen that the HD-SIC outperforms the fICIC for strong ICI, which inspires us to investigate the SIC-based fICIC in future work, where the FD PBS forwards the listened ICI to further enhance the ICI at the PUE so as to extend the feasible region of SIC.

\begin{figure}
\centering
\includegraphics[width=0.46\textwidth]{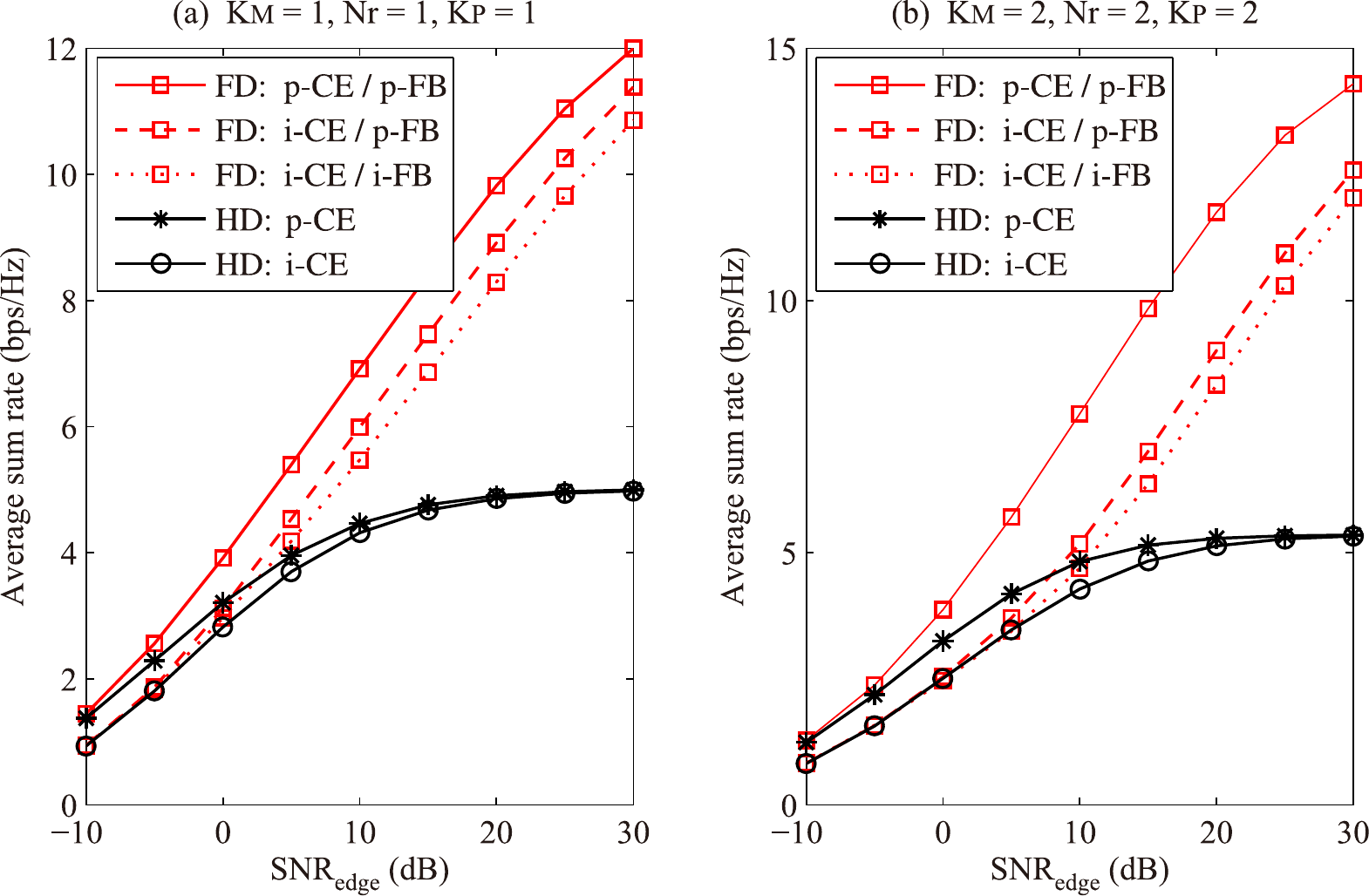}
\caption{Average sum rate of PUEs in Cell 1 versus $\text{SNR}_{\text{edge}}$, where $\bs_1$ is uniformly placed within the area suffering from strong ICI with $d_{P1}\sim U([50, 250])$~$m$, $\text{SIR}_{\text{self}} = 110$~dB, $N_r = 1, 2$, $K_M = 1, 2$, and $K_P = 1, 2$. In the legends, ``p-CE'' and ``i-CE'' denote perfect and imperfect channel estimation, respectively, and ``p-FB'' and ``i-FB'' denote perfect and imperfect feedback, respectively.}\label{F:CE}
\end{figure}

Finally, in Fig.~\ref{F:CE} the performance of the fICIC under imperfect self-interference cancellation, imperfect channel estimation and practical channel feedback is evaluated, where only Cell~1 is considered and $\bs_1$ is randomly placed within the area experiencing strong ICI, specifically with $d_{P1}$ following uniform distribution between $[50, 250]$~$m$. As discussed before, the channel $\bar{\mathbf{H}}_{MP}$ can be directly estimated at the PBS, the channel $\mathbf{h}_{Pk}$ can be obtained at the PBS by estimating the uplink channel based on channel reciprocity, and the channel $\bar{\mathbf{h}}_{Mk}$ can be first estimated at the $\ms_k$ and then fed back to the PBS. In simulations, we employ linear minimum mean-squared error estimator to estimate $\bar{\mathbf{H}}_{MP}$, $\mathbf{h}_{Pk}$ and $\bar{\mathbf{h}}_{Mk}$, and use analog feedback~\cite{Marzetta2006} to send back the estimate of $\bar{\mathbf{h}}_{Mk}$ to the PBS , where the transmit power of $\ms_k$ is set as $23$~dBm.
Figs.~\ref{F:CE}(a) and \ref{F:CE}(b) show the results in single-user and multi-user cases, respectively. In both cases we can see the performance degradation of the fICIC caused by imperfect channel estimation and imperfect feedback. However, compared with the HD scheme, significant performance gain achieved by the fICIC can be still observed, even when imperfect self-interference cancellation, imperfect channel estimation and practical analog feedback are taken into account.

\subsection{Wideband Single-user Case}
In this subsection we evaluate the performance of the fICIC in wideband systems. We consider a LTE system with $5$~MHz bandwidth, where the sampling interval is $T_s = 0.13$~$\mu$s~\cite{TS36.211}. The small-scale channels are generated based on WINNER II clustered delay line model~\cite{WinnerII}. Specifically, the channels from the MBS to the PBS, $\mathbf{H}_{MP}$, use the typical urban macro-cell line of sight (LoS) model considering that the receive antennas of the PBS can be mounted outside a building as discussed before, the channels from the MBS to PUEs, $\mathbf{h}_{Mk}$, use the typical urban macro-cell non-LoS model, and the channels from the PBS to PUEs, $\mathbf{h}_{Pk}$, use the typical urban micro-cell NLoS model. After sampling the multipath channels with the considered bandwidth, we obtain the maximal delay spread of $\mathbf{H}_{MP}$ and $\mathbf{h}_{Pk}$ as three and six samples, respectively. We consider the processing delay of the FD PBS as $4$ samples, i.e., $\tau = 0.52\ \mu$s. Since the CP of the LTE system is $4.7$~$\mu$s~\cite{TS36.211}, i.e., $36$ samples, we can obtain the maximal order of the FIR forwarding precoder $\mathbf{W}_f(t)$ as $23$, i.e., $L\leq23$. Note that the maximum value of $L$ can be even larger if the extended CP with $16.7$~$\mu$s is considered.

\begin{figure}
\centering
\includegraphics[width=0.46\textwidth]{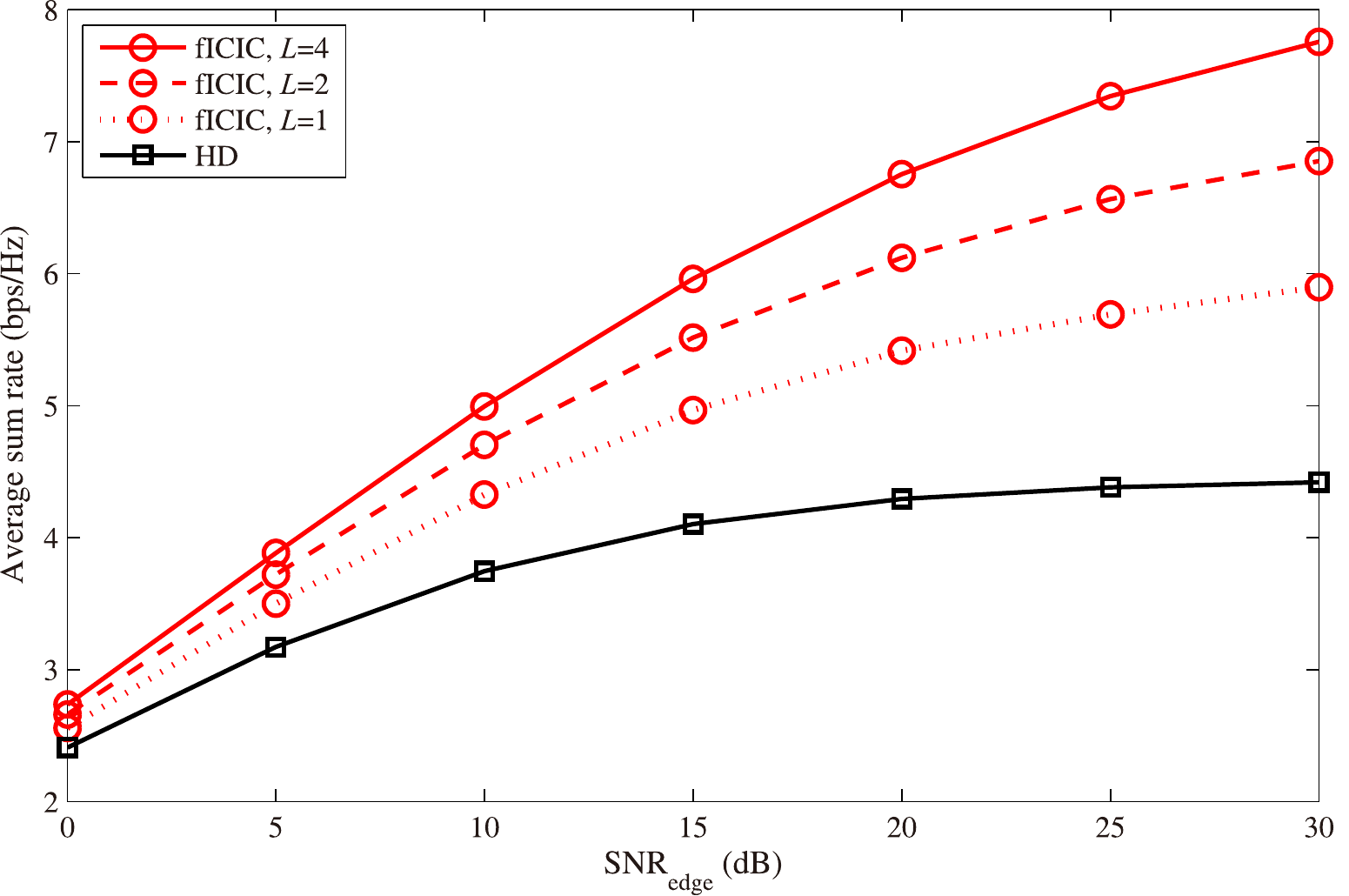}
\caption{Average sum rate of the PUE over $N=25$ subcarriers in Cell 1 versus $\text{SNR}_{\text{edge}}$, where $\text{SIR}_{\text{self}} = 110$~dB, $K_M = 1$, and $K_P = 1$.}\label{F:WB}
\end{figure}

Considering that the complexity of solving problem (\ref{E:problem-wb}) increases rapidly with the number of subcarriers and also considering the channel correlation in frequency domain, in the simulations we treat each resource block (RB) as a subcarrier and then $N=25$ subcarriers are simulated corresponding to the total $25$ RBs of the $5$~MHz LTE system~\cite{TS36.211}. In addition, we select the order of $\mathbf{W}_f$ as $L=1,2,4$ to speed up the simulations. In Fig.~\ref{F:WB}, the average sum rate of Cell 1 with $d_{P1}=60~m$ achieved by the fICIC is depicted, where imperfect self-interference cancellation, imperfect channel estimation and practical channel feedback as considered in Fig.~\ref{F:CE} are taken into account. The compared HD scheme is obtained from problem (\ref{E:problem-wb}) by setting $\mathbf{W}_f$ as zeros.

We can see from Fig.~\ref{F:WB} that the performance of the wideband fICIC improves with the increase of $L$, and an evident performance gain over the HD scheme can be observed when $L=4$. More significant performance gain can be expected when larger $L$ is used for the fICIC, to achieve which more efficient low-complexity algorithms to problem (\ref{E:problem-wb}) need to be studied in future work.

\section{Conclusions}
In this paper we proposed to eliminate the cross-tier ICI in HetNets by using FD technique at the PBS. We derived a FD assisted ICI cancellation (fICIC) scheme, with which the ICI can be mitigated by merely designing the precoders at PBSs without relying on the participation of the MBS. We first investigated the narrowband single-user case to gain some insight into the behavior of the fICIC, where we found closed-from solution of the optimal fICIC, and analyzed its asymptotical performance in ICI-dominated scenario. We then studied the general narrowband multi-user case, and devised a low-complexity algorithm to find the optimal fICIC scheme that maximizes the downlink sum rate of PUEs subject to the fairness constraint. Finally, we generalized the fICIC to wideband systems and discussed the practical issues regarding the application of the~fICIC. Simulations validated the analytical results. Compared with the traditional HD scheme, the fICIC exhibits significant performance gain even when imperfect self-interference cancellation and imperfect channel information are taken into account. By combining the fICIC with eICIC or CoMP-CB, the ICI with various levels can be effectively eliminated.

\numberwithin{equation}{section}
\appendices

\section{Derivation of Transmit Power $P_{out}$} \label{A:Pout}
By substituting (\ref{E:yp}) into (\ref{E:x2}), we can obtain
\begin{align} \label{E:xp-new}
  &\mathbf{x}_p[t] =\mathbf{W}_f \big(\bar{\mathbf{H}}_{MP}^H \mathbf{s}_M[t\!-\!\tau] - \mathbf{E}_{PP}^H[t\!-\!\tau]\mathbf{x}_p[t\!-\!\tau]+\\
  & \mathbf{H}_{PP}^H \mathbf{z}_x[t\!-\!\tau]+ \mathbf{n}_p[t\!-\!\tau] + \mathbf{z}_y[t\!-\!\tau]\big)+ \textstyle{\sum_{k=1}^{K_P}} \mathbf{w}_{d,k} s_{p,k}[t].\nonumber
\end{align}

Since the terms at the right-hand side of (\ref{E:xp-new}) are independent, we can obtain $P_{out}$ with~(\ref{E:Pout-new}) by taking the expectations over each term, yielding
\begin{align}
  P_{out}& = \mathtt{tr}\left(\mathbf{W}_f\bar{\mathbf{H}}_{MP}^H\bar{\mathbf{H}}_{MP}\mathbf{W}_f^H\right) +   \sigma_n^2\mathtt{tr}\left(\mathbf{W}_f\mathbf{W}_f^H\right)\nonumber\\
  & + \textstyle{\sum_{k=1}^{K_P}} \|\mathbf{w}_{d,k}\|^2 + \mathtt{tr}\left(\mathbf{W}_f\boldsymbol{\Sigma}_1\mathbf{W}_f^H\right) \nonumber\\
  &+ \mathtt{tr}\left(\mathbf{W}_f\boldsymbol{\Sigma}_2\mathbf{W}_f^H\right) +\mathtt{tr}\left(\mathbf{W}_f\boldsymbol{\Sigma}_3\mathbf{W}_f^H\right),\label{E:Pout-0}
\end{align}
where $\boldsymbol{\Sigma}_1 \triangleq \mathcal{E}_{\mathbf{x}_p, \mathbf{E}_{PP},\mathbf{H}_{PP}}\{\mathbf{E}_{PP}^H[t\!-\!\tau]\mathbf{x}_p[t\!-\!\tau]\mathbf{x}_p^H[t\!-\!\tau]\mathbf{E}_{PP}[t\!-\!\tau]\}$, $\boldsymbol{\Sigma}_2 \triangleq \mathcal{E}_{\mathbf{H}_{PP}}\{\mathbf{H}_{PP}^H \mu_x\mathtt{diag}(\boldsymbol{\Phi}_x) \mathbf{H}_{PP}\}$, and $\boldsymbol{\Sigma}_3\triangleq\mathcal{E}_{\mathbf{H}_{PP}}\{\mu_y\mathtt{diag}(\boldsymbol{\Phi}_y)\}$, which can be derived as follows.

Denoting $\mathbf{x}_p = [x_{p1},\dots,x_{pN_t}]^T$, we can rewrite $\boldsymbol{\Sigma}_1$~as
\begin{align} \label{E:Sigma-1}
  &\boldsymbol{\Sigma}_1= \mathcal{E}_{\mathbf{x}_p, \mathbf{E}_{PP},\mathbf{H}_{PP}}\Big\{\textstyle{\sum_{i=1}^{N_t}} |x_{pi}[t-\tau]|^2 \mathbf{e}_{PP,i}\mathbf{e}_{PP,i}^H\Big\} \nonumber\\
  &\stackrel{(a)}{=} \mathcal{E}_{\mathbf{H}_{PP}}\Big\{\textstyle{\sum_{i=1}^{N_t}} [\boldsymbol{\Phi}_x]_{ii} \tilde{\boldsymbol{\Phi}}_{e,i}\Big\}\nonumber\\
  &\stackrel{(b)}{=} \mathcal{E}_{\mathbf{H}_{PP}}\Big\{\textstyle{\sum_{i=1}^{N_t}} [\boldsymbol{\Phi}_x]_{ii} \Big(\mathbf{H}_{PP}^H \textstyle{\sum_{j=1}^{N_t}}|c_{ij}|^2\mu_x\mathtt{diag}(\mathbf{c}_j\mathbf{c}_j^H) \mathbf{H}_{PP} \nonumber \\
  &\quad  + \frac{1}{P_{tr}}(1+\mu_y)\sigma_n^2\mathbf{I}_{N_r} + \textstyle{\sum_{j=1}^{N_t}} |c_{ij}|^2 \mu_y\cdot\nonumber\\
  &\quad \mathtt{diag}\big(\mathbf{H}_{PP}^H \big(\mathbf{c}_j\mathbf{c}_j^H+\mu_x\mathtt{diag}(\mathbf{c}_j\mathbf{c}_j^H)\big)\mathbf{H}_{PP}\big)  \Big) \Big\} \nonumber\\
  & \stackrel{(c)}{=} \textstyle{\sum_{i=1}^{N_t}} [\boldsymbol{\Phi}_x]_{ii} \Big(\textstyle{\sum_{j=1}^{N_t}} |c_{ij}^2|\mu_x \bar{\alpha}_{PP} \mathtt{tr}\big(\mathtt{diag}(\mathbf{c}_j\mathbf{c}_j^H)\big)\mathbf{I}_{N_r} +\nonumber\\
  &\quad \frac{1}{P_{tr}}(1+\mu_y) \sigma_n^2\mathbf{I}_{N_r} + \textstyle{\sum_{j=1}^{N_t}} |c_{ij}|^2\mu_y \bar{\alpha}_{PP}\big(\mathtt{tr}(\mathbf{c}_j\mathbf{c}_j^H)\mathbf{I}_{N_r} +\nonumber\\
  &\quad
  \mu_x\mathtt{tr}\big(\mathtt{diag}(\mathbf{c}_j\mathbf{c}_j^H)\mathbf{I}_{N_r}\big)\big)  \Big)\nonumber\\
  &\stackrel{(d)}{=} \big(\frac{(1\!+\!\mu_y)\sigma_n^2}{P_{tr}} \!+\! \bar{\alpha}_{PP}(\mu_x\!+\!\mu_y\!+\!\mu_x\mu_y)\big) \mathtt{tr}(\boldsymbol{\Phi}_x) \mathbf{I}_{N_r},
\end{align}
where the expectations over $\mathbf{x}_p$ and $\mathbf{E}_{PP}$ are taken in step~$(a)$, $[\boldsymbol{\Phi}_x]_{ii}$ denotes the $i$-th diagonal element of $\boldsymbol{\Phi}_x$, step~$(b)$ comes from (\ref{E:Epp-cov}), step~$(c)$ takes the expectation over $\mathbf{H}_{PP}$ considering $\mathtt{vec}(\mathbf{H}_{PP})\sim\mathcal{CN}(\bar{\mathbf{0}}_{N_rN_t}, \bar{\alpha}_{PP}\mathbf{I}_{N_rN_t})$, and step~$(d)$ comes from the fact that $\mathbf{CC}^H = \mathbf{I}_{N_t}$ such that $\mathtt{tr}(\mathbf{c}_i\mathbf{c}_i^H) = \mathtt{tr}(\mathtt{diag}(\mathbf{c}_i\mathbf{c}_i^H)) = \sum_{j=1}^{N_t} |c_{ij}|^2 = 1$.

The term $\boldsymbol{\Sigma}_2$ can be obtained as
\begin{equation}  \label{E:Sigma-2}
    \boldsymbol{\Sigma}_2 = \mathcal{E}_{\mathbf{H}_{PP}}\{\mathbf{H}_{PP}^H \mu_x\mathtt{diag}(\boldsymbol{\Phi}_x) \mathbf{H}_{PP}\} = \bar{\alpha}_{PP}\mu_x\mathtt{tr}(\boldsymbol{\Phi}_x) \mathbf{I}_{N_r}.
\end{equation}

With (\ref{E:Phi-y}), the term $\boldsymbol{\Sigma}_3$ can be obtained as
\begin{align}  \label{E:Sigma-3}
    & \boldsymbol{\Sigma}_3 =\mathcal{E}_{\mathbf{H}_{PP}}\{\mu_y\mathtt{diag}(\boldsymbol{\Phi}_y)\} \nonumber\\
    & = \mathcal{E}_{\mathbf{H}_{PP}}\Big\{\mu_y\mathtt{diag}\big( \bar{\mathbf{H}}_{MP}^H \bar{\mathbf{H}}_{MP} +\nonumber\\
    &\quad \mathbf{H}_{PP}^H (\boldsymbol{\Phi}_x+\mu_x\mathtt{diag}(\boldsymbol{\Phi}_x)) \mathbf{H}_{PP} + \sigma_n^2\mathbf{I}_{N_r} \big)\Big\}\\
    & = \mu_y\Big(\mathtt{diag}(\bar{\mathbf{H}}_{MP}^H \bar{\mathbf{H}}_{MP}) \!+\! \bar{\alpha}_{PP}(1\!+\!\mu_x)\mathtt{tr}(\boldsymbol{\Phi}_x)\mathbf{I}_{N_r} \!+\! \sigma_n^2\mathbf{I}_{N_r}  \Big).\nonumber
\end{align}

Substituting (\ref{E:Sigma-1}), (\ref{E:Sigma-2}) and (\ref{E:Sigma-3}) into (\ref{E:Pout-0}) and noting that $\mathtt{tr}(\boldsymbol{\Phi}_x) = P_{out}$, we can obtain
\begin{align}\label{E:Pout-final}
  &P_{out}
  %& = \mathtt{tr}\left(\mathbf{W}_f\bar{\mathbf{H}}_{MP}^H\bar{\mathbf{H}}_{MP}\mathbf{W}_f^H\right) +   \sigma_n^2\mathtt{tr}\left(\mathbf{W}_f\mathbf{W}_f^H\right) + \sum_{k=1}^{K_P} \|\mathbf{w}_{d,k}\|^2 \label{E:Pout-f1}\\
  %& + \bar{\alpha}_{PP}(1+2\mu_x+2\sigma_n^2) \mathtt{tr}(\boldsymbol{\Phi}_x) \mathtt{tr}(\mathbf{W}_f\mathbf{W}_f^H)  \label{E:Pout-f2}\\
  %& + \bar{\alpha}_{PP}\mu_x\mathtt{tr}(\boldsymbol{\Phi}_x) \mathtt{tr}(\mathbf{W}_f\mathbf{W}_f^H)   \label{E:Pout-f3}\\
  %& + \mu_y  \mathtt{tr}\big(\mathbf{W}_f \mathtt{diag}(\bar{\mathbf{H}}_{MP}^H \bar{\mathbf{H}}_{MP}) \mathbf{W}_f^H\big) + \mu_y\big(\bar{\alpha}_{PP}(1+\mu_x)\mathtt{tr}(\boldsymbol{\Phi}_x) + \sigma_n^2 \big) \mathtt{tr}(\mathbf{W}_f\mathbf{W}_f^H) ,\label{E:Pout-f4}\\
  %= & \mathtt{tr}\left(\mathbf{W}_f\big(\bar{\mathbf{H}}_{MP}^H\bar{\mathbf{H}}_{MP} + \mu_y \mathtt{diag}(\bar{\mathbf{H}}_{MP}^H \bar{\mathbf{H}}_{MP})\big)\mathbf{W}_f^H\right) +
  %\sum_{k=1}^{K_P} \|\mathbf{w}_{d,k}\|^2 + \nonumber\\
  %& \Big(\sigma_n^2 + \big(\frac{1}{P_{tr}}(1+\mu_y)\sigma_n^2 + \bar{\alpha}_{PP}(\mu_x+\mu_y+\mu_x\mu_y)\big)P_{out}
  %+ \bar{\alpha}_{PP}\mu_x P_{out} + \nonumber\\
  %&\mu_y\big(\bar{\alpha}_{PP}(1+\mu_x) P_{out} + \sigma_n^2 \big)\Big) \mathtt{tr}(\mathbf{W}_f\mathbf{W}_f^H) \nonumber\\
  \!=\! \mathtt{tr}\left(\mathbf{W}_f\big(\bar{\mathbf{H}}_{MP}^H\bar{\mathbf{H}}_{MP} \!+\! \mu_y \mathtt{diag}(\bar{\mathbf{H}}_{MP}^H \bar{\mathbf{H}}_{MP})\big)\mathbf{W}_f^H\right) \!+\! \nonumber\\
  & \ \  \textstyle{\sum_{k=1}^{K_P}} \|\mathbf{w}_{d,k}\|^2 \!+\!\Big((1+\mu_y)\sigma_n^2 \!+\! \nonumber\\
  & \ \ \big({\frac{(1\!+\!\mu_y)\sigma_n^2}{P_{tr}}} \!+\! 2\bar{\alpha}_{PP}(\mu_x\!+\!\mu_y\!+\!\mu_x\mu_y)) \big) P_{out}\Big) \mathtt{tr}(\mathbf{W}_f\mathbf{W}_f^H) \nonumber\\
  &\approx \mathtt{tr}\left(\mathbf{W}_f\bar{\mathbf{H}}_{MP}^H\bar{\mathbf{H}}_{MP}\mathbf{W}_f^H\right) +
  \textstyle{\sum_{k=1}^{K_P}} \|\mathbf{w}_{d,k}\|^2 +\nonumber\\
   &\ \ \big(\sigma_n^2  + (\frac{\sigma_n^2}{P_{tr}}+2\bar{\alpha}_{PP}(\mu_x+\mu_y)) P_{out} \big) \mathtt{tr}(\mathbf{W}_f\mathbf{W}_f^H),
\end{align}
where the approximation follows from $\mu_x\ll 1$ and $\mu_y\ll 1$ as in~\cite{Day2012}.

\section{Proof of Lemma 1} \label{S:prooflemma2}
To prove this lemma, we rewrite problem (\ref{E:problem}) by expressing $\mathbf{W}_f$ with its vectorization, denoted by $\bar{\mathbf{w}}_f = \mathtt{vec}(\mathbf{W}_f)$, as follows.
\begin{subequations} \label{E:problem1-vec}
    \begin{align}
        &\underset{\bar{\mathbf{w}}_f, \mathbf{w}_{d,k}}{\max}\ \frac{|\mathbf{h}_{Pk}^H\mathbf{w}_{d,k}|^2}{\Omega'_k } \label{E:Objective-vec}\\
         &s.t.\ \|\big(\bar{\mathbf{h}}_{MP}^T\!\otimes\!\mathbf{I}_{N_t}\big)\bar{\mathbf{w}}_f\|^2 \!+\! (P_0\sigma_e^2 \!+\! \sigma_n^2)\|\bar{\mathbf{w}}_f\|^2 \!+\! \|\mathbf{w}_{d,k}\|^2 \!\leq\! P_0, \label{E:Constraint-vec}
    \end{align}
\end{subequations}
where $\Omega'_k \triangleq |\bar{h}_{Mk}^\ast +  e^{-j\phi} \big(\bar{\mathbf{h}}_{MP}^T \otimes\mathbf{h}_{Pk}^H \big) \bar{\mathbf{w}}_f|^2  + \|\big(\mathbf{I}_{N_r}\otimes \mathbf{h}_{Pk}^H\big) \bar{\mathbf{w}}_f\|^2(P_0\sigma_e^2+\sigma_n^2) + \sigma_n^2$, and  the property $\mathtt{vec}(\mathbf{AXB}) = \left(\mathbf{B}^T\otimes\mathbf{A}\right)\mathtt{vec}(\mathbf{X})$ is used.

Based on the KKT condition, we can obtain the optimal solution of $\bar{\mathbf{w}}_{f}$ as
\begin{align} \label{E:optwf}
   &\bar{\mathbf{w}}_f^\star = - \bar{h}_{Mk}^\ast e^{j\phi}\Big(\frac{\mathtt{SINR}_{k,FD}}{\Delta_1} \big(\bar{\mathbf{h}}_{MP}^T \otimes \mathbf{h}_{Pk}^H \big)^H\big(\bar{\mathbf{h}}_{MP}^T \otimes \mathbf{h}_{Pk}^H\big) \nonumber\\
   &\!+\! \frac{\mathtt{SINR}_{k,FD}\Delta_2}{\Delta_1} \!\big(\! \mathbf{I}_{N_r}\! \otimes\! \mathbf{h}_{Pk}^H\!\big)^H\big(\mathbf{I}_{N_r}\!\otimes\! \mathbf{h}_{Pk}^H\big)\!+\! \lambda \big(\bar{\mathbf{h}}_{MP}^T\!\otimes\!\mathbf{I}_{N_t}\big)^H\cdot\nonumber\\
   &   \big(\bar{\mathbf{h}}_{MP}^T \otimes\mathbf{I}_{N_t}\big) + \lambda\Delta_2 \mathbf{I}_{N_tN_r}\Big)^{-1} \big(\bar{\mathbf{h}}_{MP}^T \otimes\mathbf{h}_{Pk}^H\big)^H,
\end{align}
where $\Delta_1 =  |\bar{h}_{Mk}^\ast +  e^{-j\phi} \left(\bar{\mathbf{h}}_{MP}^T \otimes \mathbf{h}_{Pk}^H\right) \bar{\mathbf{w}}_f^\star|^2  + \|\left(\mathbf{I}_{N_r}\otimes \mathbf{h}_{Pk}^H\right) \bar{\mathbf{w}}_f^\star\|^2(P_0\sigma_e^2+\sigma_n^2) + \sigma_n^2$, $\Delta_2 = P_0\sigma_e^2+\sigma_n^2$, and $\lambda$ is the lagrangian multiplier.

By applying the properties of Kronecker product and after some regular manipulations, we can simplify (\ref{E:optwf}) as
\begin{align} \label{E:optwf-sim}
   \bar{\mathbf{w}}_f^\star %= & - \bar{h}_{Mk}^\ast e^{j\phi}\left(\frac{\mathtt{SINR}_{k,FD}}{\Delta_1} \left((\bar{\mathbf{h}}_{MP}^H)^T \otimes \mathbf{h}_{Pk} \right)\left( \bar{\mathbf{h}}_{MP}^T\otimes\mathbf{h}_{Pk}^H\right) + \frac{\mathtt{SINR}_{k,FD}\Delta_2}{\Delta_1} \left(\mathbf{I} \otimes \mathbf{h}_{Pk} \right)\left(\mathbf{I} \otimes \mathbf{h}_{Pk}^H\right) \right. \nonumber\\
   = & - \bar{h}_{Mk}^\ast e^{j\phi}\left(\left((\bar{\mathbf{h}}_{MP}\bar{\mathbf{h}}_{MP}^H)^T +  \Delta_2\mathbf{I}_{N_r}\right)^{-1} (\bar{\mathbf{h}}_{MP}^H)^T\right)\nonumber\\
   &\otimes \left(\left(\frac{\mathtt{SINR}_{k,FD}}{\Delta_1}\mathbf{h}_{Pk}\mathbf{h}_{Pk}^H + \lambda\mathbf{I}_{N_t} \right)^{-1}\mathbf{h}_{Pk}\right).
\end{align}

Then based on the matrix inversion lemma, we can further rewrite (\ref{E:optwf-sim}) as
\begin{align} \label{E:optwf2}
\bar{\mathbf{w}}_{f}^\star &= \frac{ - \bar{h}_{Mk}^\ast e^{j\phi}\Delta_1(\bar{\mathbf{h}}_{MP}^H)^T\otimes \mathbf{h}_{Pk}}{\left(\mathtt{SINR}_{k,FD}\|\mathbf{h}_{Pk}\|^2 + \lambda\Delta_1\right)\left( \|\mathbf{h}_{MP}\|^2 + \Delta_2\right)} \nonumber\\
&\triangleq  - \bar{h}_{Mk}^\ast e^{j\phi}\cdot \beta \cdot (\bar{\mathbf{h}}_{MP}^H)^T\otimes\mathbf{h}_{Pk},
\end{align}
where $\beta = \frac{\Delta_1}{\left(\mathtt{SINR}_{k,FD}\|\mathbf{h}_{Pk}\|^2 + \lambda\Delta_1\right)\left( \|\mathbf{h}_{MP}\|^2 + \Delta_2\right)}$ is a positive scalar.

According to (\ref{E:optwf2}), we can recover the optimal $\mathbf{W}_f^\star$ from $\bar{\mathbf{w}}_f^\star$ as (\ref{E:w1}), which is of rank 1.

\section{Proof of Proposition 1} \label{S:prooftheorem1}
By defining $\nu \triangleq |\bar{h}_{Mk}|\|\mathbf{h}_{Pk}\|\|\bar{\mathbf{h}}_{MP}\| \beta = \frac{C}{2}\beta$, problem (\ref{E:problem1}) can be rewritten as
\begin{subequations} \label{E:problem1-A}
    \begin{align}
        \underset{\nu}{\max}\ & f(\nu) \triangleq \frac{A-B\nu^2}{B\nu^2 - C\nu + D}  \label{E:Objective1-A}\\
         s.\ t.\ & \nu^2 \leq \frac{A}{B}\label{E:Constraint1-A}\\
         \ & \nu \geq 0, \label{E:Constraint2-A}
    \end{align}
\end{subequations}
where the parameters $A$, $B$, $C$ and $D$ are defined in Proposition~1.

Since the numerator of the objective function (\ref{E:Objective1-A}) is concave, the denominator is
convex, and both are differential, we know that the objective function of problem (\ref{E:problem1-A}) is quasi-concave~\cite{boyd2009convex}. This suggests that the global optimal solution to problem (\ref{E:problem1-A}), denoted by $\bar \nu$, needs to satisfy the following KKT conditions,
\begin{subequations} \label{E:KKT-A}
    \begin{align}
         -\nabla f(\bar{\nu}) + 2u\bar{\nu} - v &= 0  \label{E:KKT-A1}\\
          u(\bar{\nu}^2 - \frac{A}{B}) = 0,\ u &\geq 0 \label{E:KKT-A2}\\
          v \bar{\nu} = 0,\ v &\geq 0, \label{E:KKT-A3}
    \end{align}
\end{subequations}
where $u$ and $v$ are the lagrangian multipliers.

We next show that $u = 0$ and $v = 0$ for problem (\ref{E:problem1-A}). First, we find that $\bar{\nu}^2 < \frac{A}{B}$ must hold for the constraint (\ref{E:Constraint1-A}), otherwise, when $\bar{\nu}^2 = \frac{A}{B}$, the objective function will be zero that is obviously non-optimal. Therefore, from (\ref{E:KKT-A2}) we have $u = 0$. Second, the objective function satisfies $\nabla f(0) > 0$, which means that we can always find a small $\epsilon > 0$ making $f(\epsilon) > f(0)$, i.e., $\bar{\nu} > 0$ must hold. Therefore, based on (\ref{E:KKT-A3}) we have $v = 0$.

Then the condition (\ref{E:KKT-A1}) for $\bar{\nu}$ can be simplified as
\begin{equation}\label{E:Largrange}
  2B\bar{\nu} - \frac{(B\bar{\nu}^2-A)(2B\bar{\nu}-C)}{B\bar{\nu}^2 - C\bar{\nu} + D} = 0.
\end{equation}

From (\ref{E:Largrange}), we can derive the optimal value of the objective function as
\begin{equation}
  f(\bar{\nu}) =\frac{2B\bar{\nu}}{C-2B\bar{\nu}} = \frac{1}{\frac{C}{2B\bar{\nu}} -1}.
\end{equation}

The condition (\ref{E:Largrange}) can be further expressed as a quadratic equation with respect to $\bar{\nu}$, which~is
\begin{equation}\label{E:equation}
  g(\bar{\nu}) \triangleq BC\bar{\nu}^2 - 2B(A+D)\bar{\nu} + AC = 0,
\end{equation}
whose two solutions can be obtained as
\begin{align}\label{E:two-solutions}
  \bar{\nu} &= \frac{A+D\pm\sqrt{(A+D)^2 - \frac{AC^2}{B}}}{C}.
\end{align}

The feasibility of the two solutions is examined as follows. Recalling that $0\leq \bar{\nu}\leq \sqrt{\frac{A}{B}}$ from (\ref{E:Constraint1-A}) and (\ref{E:Constraint2-A}), we can see that $g(0) = AC > 0$ and $g(\small{\sqrt{\frac{A}{B}}})<0$ because
\begin{align}
  & g(\textstyle{\sqrt{\frac{A}{B}}})= 2AC - 2(A+D)\sqrt{AB} \nonumber\\
  & < 4P_0|\bar{h}_{Mk}|\|\bar{\mathbf{h}}_{MP}\|\|\mathbf{h}_{Pk}\|^3
  - 2(P_0\|\mathbf{h}_{Pk}\|^2 + |\bar{h}_{Mk}|^2)\cdot\nonumber\\
  &\qquad\sqrt{P_0\|\bar{\mathbf{h}}_{MP}\|^2\|\mathbf{h}_{Pk}\|^4}\nonumber\\
  & =  -2\sqrt{P_0}\|\bar{\mathbf{h}}_{MP}\|\|{\mathbf{h}}_{Pk}\|^2(|\bar{h}_{Mk}| - \sqrt{P_0}\|\mathbf{h}_{Pk}\|)^2,
\end{align}
where the inequality is obtained by setting $\sigma_e^2 = 0$ and $\sigma_n^2 = 0$ in the definitions of $B$ and $D$.

The results indicate that equation (\ref{E:equation}) has one and only one solution within $[0, \sqrt{\frac{A}{B}}]$, which is the smaller one in (\ref{E:two-solutions}). Finally, recalling that $\nu = \frac{C}{2}\beta$, we can obtain the optimal $\beta$ as shown after~(\ref{E:maxSINR}).

\section{Proof of Proposition 2} \label{S:prooflemma4}
By defining $\mathbf{A} = \mathbf{I}_{N_t} +\sum_{j\neq k} \bar{\lambda}_j \mathbf{h}_{Pj}\mathbf{h}_{Pj}^H$ and $\mathbf{B} = \mathbf{I}_{N_t} - \frac{\bar{\lambda}_k}{\gamma_k} \mathbf{A}^{-\frac{1}{2}} \mathbf{h}_{Pk} \mathbf{h}_{Pk}^H\mathbf{A}^{-\frac{H}{2}}$, we can express the left-hand side of (\ref{E:con-lambda}) as $\mathbf{A}^{\frac{1}{2}}\mathbf{B}\mathbf{A}^{\frac{H}{2}}$. Since $\mathbf{A}$ is positive definite and $\mathbf{B}$ is Hermitian, it is not difficult to show that $\mathbf{A}^{\frac{1}{2}}\mathbf{B}\mathbf{A}^{\frac{H}{2}}\succeq \mathbf{0}_{N_t}$ and $\mathbf{B}\succeq \mathbf{0}_{N_t}$ are equivalent to each other. Then, the semi-definite positive constraints (\ref{E:con-lambda}) can be equivalently expressed as
\begin{align} \label{E:semi-definite}
  & \mathbf{I}_{N_t} - \frac{\bar{\lambda}_k}{\gamma_k}\big(\mathbf{I}_{N_t} +\sum_{j\neq k} \bar{\lambda}_j \mathbf{h}_{Pj}\mathbf{h}_{Pj}^H\big)^{-\frac{1}{2}}  \mathbf{h}_{Pk} \mathbf{h}_{Pk}^H\big(\mathbf{I}_{N_t} \nonumber\\
  & \quad\qquad+\sum_{j\neq k} \bar{\lambda}_j \mathbf{h}_{Pj}\mathbf{h}_{Pj}^H\big)^{-\frac{H}{2}} \succeq \mathbf{0}_{N_t}, \forall k.
\end{align}

The positive semi-definite constraint in (\ref{E:semi-definite}) means that the minimal eigenvalue of the matrix in the left-hand side should be non-negative. Note that the term $\left(\mathbf{I}_{N_t} +\sum_{j\neq k} \bar{\lambda}_j \mathbf{h}_{Pj}\mathbf{h}_{Pj}^H\right)^{-\frac{1}{2}}$ $\mathbf{h}_{Pk} \mathbf{h}_{Pk}^H\left(\mathbf{I}_{N_t} +\sum_{j\neq k} \bar{\lambda}_j \mathbf{h}_{Pj}\mathbf{h}_{Pj}^H\right)^{-\frac{H}{2}}$ is of rank one. It has only one positive eigenvalue, which is $\mathbf{h}_{Pk}^H\left(\mathbf{I}_{N_t} +\sum_{j\neq k} \bar{\lambda}_j \mathbf{h}_{Pj}\mathbf{h}_{Pj}^H\right)^{-1} \mathbf{h}_{Pk}$, and all other eigenvalues are zeros. Therefore, the constraint in (\ref{E:semi-definite}) can be further converted into the constraint on the minimal eigenvalue of the matrix in the left-hand side of (\ref{E:semi-definite}) as
\begin{equation}
  1 - \frac{\bar{\lambda}_k}{\gamma_k}\mathbf{h}_{Pk}^H\left(\mathbf{I}_{N_t} +\sum_{j\neq k} \bar{\lambda}_j \mathbf{h}_{Pj}\mathbf{h}_{Pj}^H\right)^{-1} \mathbf{h}_{Pk} \geq 0, \forall k.
\end{equation}
which can be further rewritten as (\ref{E:Con-semidefinite}).

\bibliography{IEEEabrv,Hanbib_D2D}

%\end{CJK*}
\end{document}